\documentclass[3p,authoryear]{elsarticle}
\usepackage{epstopdf}
\usepackage{subfigure,graphicx}
\usepackage{float}
\usepackage{amsmath, amsfonts, amssymb}
\usepackage{amsthm}
\usepackage{epsf}
\usepackage{amsfonts}
\usepackage{stmaryrd}
\usepackage{amssymb}
\usepackage{leftidx}
\usepackage{color}
\usepackage{mathtools}
\usepackage{placeins}
\usepackage{booktabs}
\usepackage{enumitem}
\usepackage{caption}
\usepackage{multirow}
\usepackage{stackengine}
\usepackage[utf8]{inputenc}
\usepackage[english]{babel}
\usepackage{color}
\usepackage{bm}

\theoremstyle{plain}
\newtheorem{theorem}{Theorem}

\newtheorem{remark}[theorem]{Remark}

\def\bfx{{\bf x}}
\def\bfy{{\bf y}}

\def\bfC{{\bf C}}

\def\bfN{{\bf N}}

\def\bfR{{\bf R}}
\def\bfS{{\bf S}}

\def\bfX{{\bf X}}
\def\bfF{{\bf F}}
\def\bfU{{\bf U}}

\def\bfe{{\bf e}}

\def\l{\lambda}

\def\e0{\varepsilon_0}

\def\s0{\sigma_0}

\long\def\symbolfootnote[#1]#2{\begingroup%
\def\thefootnote{\fnsymbol{footnote}}\footnote[#1]{#2}\endgroup}

\newcommand\sts{s_{\texttt{ts}}}

\newcommand\shs{s_{\texttt{hs}}}
\newcommand\sbs{s_{\texttt{bs}}}

\renewcommand\e{\varepsilon}

\long\def\symbolfootnote[#1]#2{\begingroup%
\def\thefootnote{\fnsymbol{footnote}}\footnote[#1]{#2}\endgroup}

\makeatletter
\renewcommand\@biblabel[1]{}

\makeatother

\begin{document}
\begin{frontmatter}

\title{The poker-chip experiments of synthetic elastomers \vspace{0.1cm}}

\author{Farhad Kamarei}
\ead{kamarei2@illinois.edu}

\author{Aditya Kumar}
\ead{aditya.kumar@ce.gatech.edu}

\author{Oscar Lopez-Pamies}
\ead{pamies@illinois.edu}

\address[illinois]{Department of Civil and Environmental Engineering, University of Illinois, Urbana--Champaign, IL 61801, USA \vspace{0.05cm}}

\address[georgia]{School of Civil and Environmental Engineering, Georgia Institute of Technology, GA 30332, USA \vspace{0.05cm}}

\vspace{0.1cm}

\begin{abstract}

\vspace{0.2cm}

In a recent study, Kumar and Lopez-Pamies (J. Mech. Phys. Solids 150: 104359, 2021) have provided a complete quantitative explanation of the famed poker-chip experiments of Gent and Lindley (Proc. R. Soc. Lond. Ser. A 249: 195--205, 1959) on natural rubber. In a nutshell, making use of the fracture theory of Kumar, Francfort, and Lopez-Pamies (J. Mech. Phys. Solids 112: 523--551, 2018), they have shown that the nucleation of cracks in poker-chip experiments in natural rubber is governed by the strength --- in particular, the hydrostatic strength --- of the rubber, while the propagation of the nucleated cracks is governed by the Griffith competition between the bulk elastic energy of the rubber and its intrinsic fracture energy. The main objective of this paper is to extend the theoretical study of the poker-chip experiment by Kumar and Lopez-Pamies to synthetic elastomers that, as opposed to natural rubber: ($i$) may feature a hydrostatic strength that is larger than their uniaxial and biaxial tensile strengths and ($ii$) do not exhibit strain-induced crystallization. A parametric study, together with direct comparisons with recent poker-chip experiments on a silicone elastomer, show that these two different material characteristics have a profound impact on where and when cracks nucleate, as well as on where and when they propagate. In conjunction with the results put forth earlier for natural rubber, the results presented in this paper provide a complete description and explanation of the poker-chip experiments of elastomers at large. As a second objective, this paper also introduces a new fully explicit constitutive prescription for the driving force that describes the material strength in the fracture theory of Kumar, Francfort, and Lopez-Pamies.

\vspace{0.2cm}

\endkeyword

\end{abstract}

\end{frontmatter}

\vspace{0.1cm}

\section{Introduction}

The famed poker-chip experiments of \cite{GL59} are widely regarded as the cradle\footnote{See, nevertheless, the works of \cite{Busse38} and \cite{Yerzley39} for earlier preliminary poker-chip experiments on natural and synthetic (neoprene) rubber.} of the investigations into nucleation on internal cracks in elastomers, a phenomenon popularly referred to as \emph{cavitation}, a term that \cite{GL57} themselves coined.

\begin{figure}[t!]
\centering
\centering\includegraphics[width=0.85\linewidth]{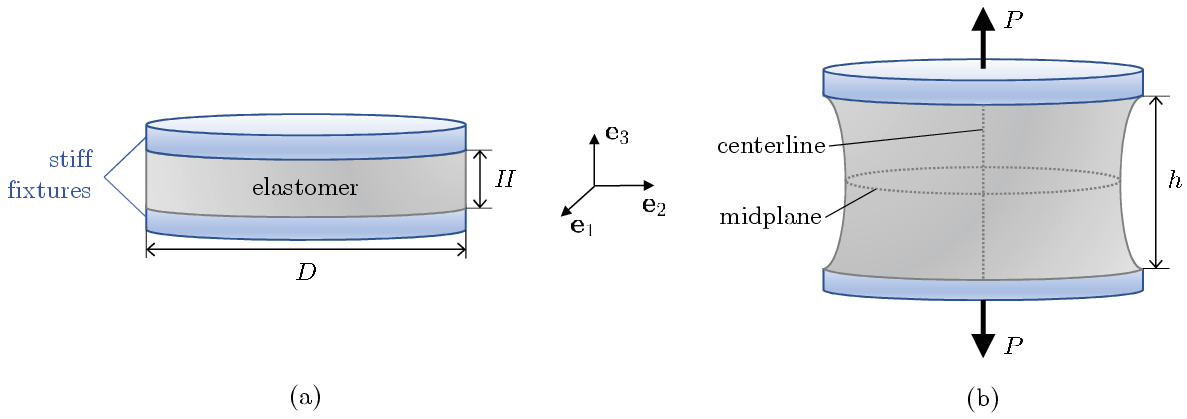}
\caption{\small Schematic of the poker-chip experimental setup in (a) the initial configuration and in (b) a deformed configuration at an applied deformation $h$ and corresponding tensile force $P$. Relative to its initial thickness $H$, the initial diameter $D$ of the elastomer disk --- the ``poker chip'' --- is typically in the range $D/H\in[2,50]$. For reference in the text, the midplane and centerline of the specimen are marked in part (b).}\label{Fig1}
\end{figure}
As schematically depicted in Fig. \ref{Fig1}, the poker-chip experiment consists in firmly bonding a thin disk of the elastomer to stiff fixtures and then pulling the fixtures apart in a testing machine. \cite{GL59} made use of specimens with numerous (in excess of twenty) diameter-to-thickness ratios $D/H$ spanning the range $D/H\in[2,50]$ and carried out experiments on eight different elastomers with a spectrum of different elastic behaviors. Specifically, they carried out experiments on seven different vulcanizates of natural rubber, four unfilled and the other three filled with different volume fractions of carbon black, as well as on one type of styrene-butadiene rubber (SBR). One of the unfilled vulcanizates (labeled \texttt{G} in their paper) was transparent, which allowed them to observe \emph{in situ} the appearance of cracks within the specimen during the loading process. The wealth of experimental observations generated by \cite{GL59} can be catalogued into the following three types:

\begin{itemize}

\item{\emph{The force-deformation response}. The primary quantitative observation reported by \cite{GL59} corresponds to plots of the normalized global force $S=4P/(\pi D^2)$ versus the normalized global deformation $\lambda=h/H$ of the poker-chip specimens up to a large force ($S=5.88$ MPa) but \emph{not} all the way until their complete failure (see Fig. 1 in their paper). These plots exhibit an initial stiffening region, followed by a plateau, followed by a secondary stiffening region before complete rupture is reached.}

\item{\emph{In situ and post mortem images of the cracks}. \cite{GL59} also reported images of the nucleated/propagated cracks in the interior of the specimens. For the transparent vulcanizate \texttt{G}, they provided an \emph{in situ} image illustrating that the first appearance of a crack occurs around the centerline of the specimen through its midplane (see Fig. 3 in their paper). For vulcanizate \texttt{D}, the unfilled rubber for which most of the reported results pertain to, they included \emph{post-mortem} images of the midplane of specimens cut open right after reaching the normalized force $S=2.75$ MPa (see Fig. 4 in their paper).

Importantly, for the thinner specimens, according to the transparent vulcanizate \texttt{G}, the sequence of events is such that a crack (or a few cracks) first appears around the centerline of the specimen. Upon further loading, more cracks continue to appear radially away until the midplane is well populated with cracks, save for a boundary layer around the free lateral boundary. For instance, for vulcanizate \texttt{D}, the post-mortem image of the midplane for a very thin specimen with diameter-to-thickness ratio $D/H=33$ shows the presence of about 100 cracks.

As the diameter-to-thickness ratio $D/H$ decreases, the post-mortem images of the midplanes show that fewer cracks get nucleated. For vulcanizate \texttt{D}, a thick specimen with diameter-to-thickness ratio $D/H=3.5$ shows only one nucleated crack, while the next thicker specimen with diameter-to-thickness ratio $D/H=2$ shows no evidence of cavitation.

Since no results for the transparent vulcanizate \texttt{G} were reported for thicker specimens, the location and the sequence of nucleation events of the various cracks and whether they propagate are unknown for those. The elongated shape of the cracks in the post-mortem images of the midplanes do suggest that they are nucleated away from the midplane, although still around the centerline, and that they propagate significantly with continuing loading.}

\item{\emph{The local-maximum stress $S^{\prime}$}. Finally, \cite{GL59} reported plots of a critical stress $S^{\prime}$ versus the poker-chip initial thickness $H$ and the elastomer Young's modulus $E$ (see Figs. 6 and 7 in their paper), which they defined as the value of the normalized force $S$ at the first local maximum in the force-deformation ($S$ \emph{vs}. $\lambda$) response of the specimens, presumed to be a proxy for the first nucleation of cracks, and therefore referred to as the ``cracking stress''.}

\end{itemize}

Despite being commonly credited with initiating the field of cavitation in elastomers, it was only in 2015 that a direct comparison\footnote{See, however, the work of \cite{Abeyaratne89} for an earlier preliminary comparison.} between a theoretical explanation and the founding poker-chip experiments of \cite{GL59} was presented in the literature by \cite{LRLP15}. That direct comparison indicated that cavitation as reported by \cite{GL59} is first and foremost a fracture process, and \emph{not} a purely elastic one as conjectured for decades\footnote{For a comprehensive account of the fascinating history of cavitation in elastomers, the reader is referred to the recent review by \cite{BCLP24}.}.

Guided by the findings of \cite{LRLP15}, and by the plethora of experimental results accumulated in the literature since the early 1900s on the nucleation and propagation of cracks within the bulk of elastomers as well as from pre-existing cracks, notches, and other boundary points \citep{Busse34,RT53,Greensmith60,Hamed2016,Chen17,Poulain17,Poulain18}, within the basic setting of negligible viscous dissipation, \citet*{KFLP18} then introduced a complete theory --- regularized, of phase-field type --- to model the nucleation and propagation of fracture in elastic brittle solids undergoing arbitrarily large quasistatic deformations. The theory can be viewed as a natural generalization of the phase-field approximation \citep{Bourdin00} of the celebrated variational theory of brittle fracture of \cite{Francfort98} --- which is nothing more than the mathematical statement of Griffith's  competition of bulk and fracture energies in its general form \citep{Griffith21} --- to account for the material strength. As such, the theory is a \emph{top-down} or \emph{macroscopic} theory of fracture that is based on three intrinsic macroscopic material properties. When specialized to elastomers, these are:
\begin{enumerate}[label=\Roman*.]

\item{the stored-energy function $$W(\bfF)$$ describing the elasticity of the elastomer for arbitrary deformation gradients $\bfF$,}

\item{the strength surface $$\mathcal{F}(\bfS)=0$$ describing the strength of the elastomer for arbitrary first Piola-Kirchhoff stress tensors $\bfS$, and}

\item{the critical energy release rate $$G_c$$ describing the intrinsic fracture energy of the elastomer, that is, the amount of energy per unit undeformed area required to create new surface in the elastomer (from an existing crack).}

\end{enumerate}
Consistent with the aforementioned plethora of experimental results amassed for over a century, the theory describes that \emph{nucleation of fracture} in a body under a uniform state of stress is governed by the strength of the elastomer, from large\footnote{``Large'' refers to large relative to the characteristic size of the underlying heterogeneities in the elastomer under investigation. By the same token, ``small'' refers to sizes that are of the same order or just moderately larger than the sizes of the heterogeneities.} pre-existing cracks is governed by the Griffith competition between the elastic energy and the intrinsic fracture energy, while under any other circumstance (e.g., from boundary points, smooth or sharp, small pre-existing cracks, or any other subregion in the body under a non-uniform state of stress) is governed by an interpolating interaction among the strength and the Griffith competition. At the same time, the theory describes that \emph{propagation of fracture} is, akin to nucleation from large pre-existing cracks, also governed by the Griffith competition between the elastic and fracture energies.

As part of a program aimed at validating its status as a complete theory of fracture by direct comparisons with experiments \emph{not} just on elastomers but on nominally elastic brittle materials at large \citep{KFLP18,KRLP18,KLP20,KBFLP20,KRLP22,KLDLP24}, \cite{KLP21} have deployed the fracture theory of \citet*{KFLP18} to simulate the seminal poker-chip experiments of \cite{GL59}. The results have served to explain in a detailed and quantitative manner all the three types of experimental observations outlined above and in so doing they have revealed that
\begin{enumerate}[label=\roman*.,font=\itshape]

\item{the nucleation of cracks in the poker-chip experiments of \cite{GL59} on natural rubber is governed by the strength --- in particular, the hydrostatic strength $\shs$ --- of the rubber, while}

\item{the propagation of the nucleated cracks is governed by the Griffith competition between the bulk elastic energy of the rubber and its intrinsic fracture energy $G_c$.}

\end{enumerate}

Now, two distinctive characteristics of natural rubber are that: ($i$) its hydrostatic strength is substantially smaller than its uniaxial and biaxial tensile strengths and ($ii$) it exhibits strain-induced crystallization.

In particular, as elaborated in Subsection 4.2 of \citep{KLP21}, the uniaxial and biaxial tensile strengths of the vulcanizate $\texttt{D}$ used by \cite{GL59}  are $s_{\texttt{ts}}=11.3\pm 3.7$ MPa and $s_{\texttt{bs}}=14.8\pm 6.2$ MPa, while its hydrostatic strength\footnote{All three strength values $s_{\texttt{ts}}$, $s_{\texttt{bs}}$, $s_{\texttt{hs}}$ referred to the nominal stresses defined by the points $\mathcal{F}({\rm diag}(s_{\texttt{ts}},0,0))=0$, $\mathcal{F}({\rm diag}(s_{\texttt{bs}},s_{\texttt{bs}},0))=0$, and $\mathcal{F}({\rm diag}(s_{\texttt{hs}},s_{\texttt{hs}},s_{\texttt{hs}}))=0$, with $s_{\texttt{ts}}, s_{\texttt{bs}}, s_{\texttt{hs}}>0$, in terms of the strength surface $\mathcal{F}(\bfS)=0$.} is $s_{\texttt{hs}}=2.9\pm 0.5$ MPa. That is, the hydrostatic strength is about four times smaller than the uniaxial and biaxial tensile strengths. This is the reason why the nucleation of cracks in the poker-chip experiments of \cite{GL59} are dominated by the (weaker) hydrostatic strength of the rubber.

Moreover, as elaborated in Subsection 4.3 of \citep{KLP21}, the strain-induced crystallization of the vulcanizate $\texttt{D}$ used by \cite{GL59} results in a very large increase in its resistance to crack growth at large stretches, one that can be approximately described in terms of an ``effective'' critical energy release rate that is \emph{not} a constant but an increasing function of deformation with values ranging from $G_c=50$ N/m when the rubber is undeformed to $G_c=10,000$ N/m when the rubber is fully stretched. This is the reason why the force-deformation ($S$ \emph{vs}. $\lambda$) response of the poker-chip experiments of \cite{GL59} exhibits a secondary stiffening region at large stretches before complete rupture is reached.

In stark contrast to natural rubber, synthetic elastomers may feature a hydrostatic strength that is larger than their uniaxial and biaxial tensile strengths and, by and large, do not exhibit strain-induced crystallization. Because of these two different material characteristics, where and when fracture nucleates and propagates in a given problem can potentially be very different for synthetic elastomers than for natural rubber.

In this context, the main objective of this paper is to extend the theoretical study by \cite{KLP21} of the poker-chip experiment on natural rubber to synthetic elastomers\footnote{Here, it is fitting to remark that the experiment that has been utilized the most to study cavitation in elastomers is, by far, the poker-chip experiment and that most of such experiments have been performed on synthetic elastomers as opposed to natural rubber; see, e.g., \citep{Lindsey67,Kakavas91,Creton10,Euchler20,LeMenn22,GuoRavi23}.}. As a second objective, this paper also introduces a new fully explicit constitutive prescription for the driving force that describes the material strength in the fracture theory of \citet*{KFLP18}.

The organization of this paper is as follows. We begin in Section \ref{Sec: strength analysis} by presenting an elementary strength analysis of the poker-chip experiment for specimens with a range of diameter-to-thickness ratios $D/H$ made of synthetic elastomers featuring different strength surfaces $\mathcal{F}(\bfS)=0$. This analysis is aimed at illustrating that the strength surface of a synthetic elastomer with a large hydrostatic strength relative to its uniaxial tensile strength --- counter to natural rubber --- can be first violated substantially away from the centerline of the specimen. In Section \ref{Sec: The theory}, we recall the fracture theory of \citet*{KFLP18} and spell out its specialization leading to the governing equations of the poker-chip experiment. We devote Subsection \ref{Sec: New ce} to introducing the new constitutive prescription for the driving force that describes the strength of the elastomer in the theory. In Subsection \ref{Sec: delta}, we then show that such a prescription can be made fully explicit. In Section \ref{Sec: 2D results}, we deploy the theory presented in Section \ref{Sec: The theory} to work out simulations of the deformation and the nucleation and propagation of fracture in poker-chip experiments for specimens with a range of diameter-to-thickness ratios $D/H$ made of elastomers featuring different strength surfaces $\mathcal{F}(\bfS)=0$ and critical energy release rates $G_c$. In Section \ref{Sec: Experiments}, we confront the predictions from the theory with the recent poker-chip experiments of \cite{GuoRavi23} on a silicone elastomer. We conclude in Section \ref{Sec: Final Comments} by summarizing the main findings of this work and by recording a number of final comments.

\section{An elementary strength analysis of the poker-chip experiment}\label{Sec: strength analysis}

\subsection{Initial configuration and kinematics}\label{Sec: Initial Conf}

Consider a poker-chip specimen of diameter $D=2.5$ cm and thickness in the range $H\in[0.0625,0.625]$ cm that in its initial (undeformed and stress-free) configuration at time $t=0$ occupies the open domain
\begin{equation*}
\Omega_0=\left\{\bfX:\sqrt{X_1^2+X_2^2}< \dfrac{D}{2},\, |X_3|< \dfrac{H}{2}\right\}
\end{equation*}
with respect to the Cartesian laboratory frame of reference $\{\bfe_i\}$ ($i=1,2,3$); see Fig. \ref{Fig1}. These specific value for $D$ and range of values for $H$ are chosen here because they are representative of typical poker-chip specimens. Note, in particular, that they span the range of diameter-to-thickness ratios
\begin{equation*}
\dfrac{D}{H}\in[4,40].
\end{equation*}
Making use of standard notation, we denote the initial boundary of the specimen by $\partial\Omega_0$ and its outward unit normal by $\bfN$.

At a later time $t\in(0,T]$, in response to the applied boundary conditions described in Subsection \ref{Sec: BCs} below, the position vector $\bfX$ of a material point in the specimen will move to a new position specified by
\begin{equation*}
\bfx = \bfy(\bfX,t),\label{y}
\end{equation*}
where $\bfy(\bfX,t)$ is a mapping from $\Omega_0$ to the current configuration $\Omega(t)$. We write the associated deformation gradient at $\bfX$ and $t$ as
\begin{equation*}
\bfF(\bfX,t)=\frac{\partial\bfy }{\partial\bfX}(\bfX,t)=\nabla \bfy(\bfX,t).
\end{equation*}

For later reference, we also spell out here the right polar decomposition of the deformation gradient in the standard form $\bfF=\bfR\bfU$, in terms of the rigid rotation tensor $\bfR$ and the right stretch tensor $\bfU$, and introduce the Cauchy-Green deformation tensor $\bfC=\bfF^T\bfF=\bfU^2$.

\subsection{Constitutive behavior of the elastomer: Elasticity, strength, and critical energy release rate}

The specimen is taken to be made of a homogeneous, isotropic, elastic brittle elastomer. Its mechanical behavior is hence characterized by three intrinsic properties: ($i$) its elasticity, ($ii$) its strength, and ($iii$) its critical energy release rate.

\begin{table}[h!]\centering
\caption{Material constants describing the elasticity, the strength, and the intrinsic fracture energy of the synthetic elastomers used in the simulations.}
\begin{tabular}{l|ccccc}
\toprule
Elasticity constants & $\mu_1$ (MPa)& $\mu_2$ (MPa) & $\alpha_1$  & $\alpha_2$ & $\kappa$ (MPa)\\
\midrule
                     & $0.0319$ & $0.0186$ & $1.391$ & $-1.021$  & $50.5$ \\
\midrule
\midrule
Strength constants   & $s_{\texttt{ts}}$ (MPa) & $s_{\texttt{hs}}$ (MPa) & $\shs/\sts$  &  & \\
\midrule
                     & $0.24$ & $0.12,0.36,0.72$ & $\dfrac{1}{2}$, $\dfrac{3}{2}$, $3$ &   &  \\
\midrule
\midrule
Critical energy release rate   & $G_c$ (N/m) &   &  &  &  \\
\midrule
                     & $10, 75, 150, 200, 500$ & &  &  & \\
\bottomrule
\end{tabular} \label{Table1}
\end{table}

\subsubsection{The elasticity}\label{Sec: Elasticity}

Precisely, we take the elastic behavior of the elastomer to be characterized by the non-Gaussian stored-energy function \citep{LP10}
\begin{equation}
W(\bfF)=\mathcal{W}(I_1,J)=\sum_{r=1}^{2}\dfrac{3^{1-\alpha_{r}}}{2\alpha_r}\mu_r\left[I_1^{\alpha_r}-3^{\alpha_r}\right]-\sum_{r=1}^{2}\mu_r\ln J+\dfrac{\kappa}{2}(J-1)^2,\label{W-LP}
\end{equation}
where
\begin{equation*}
I_1=\bfF\cdot\bfF={\rm tr}\, \bfC\qquad {\rm and}\qquad J=\det\bfF=\sqrt{\det\bfC}
\end{equation*}
stand for the first and third principal invariants of the right Cauchy-Green deformation tensor $\bfC$, while $\mu_r$, $\alpha_r$, and $\kappa$ are material constants.

It follows that the first Piola-Kirchhoff stress at any material point $\bfX\in\Omega_0$ and time $t\in[0,T]$ is given by
\begin{equation*}
\boldsymbol{\bfS}(\bfX,t)=\dfrac{\partial W}{\partial \bfF}(\bfF)=\underbrace{\left(\sum_{r=1}^{2}3^{1-\alpha_{r}}\mu_r I_1^{\alpha_r-1}\right)}_{2\mathcal{W}_{I_{1}}}\bfF+\underbrace{\left(\kappa(J-1)-\sum_{r=1}^{2}\dfrac{\mu_r}{J}\right)}_{\mathcal{W}_{J}}J\bfF^{-T},
\end{equation*}
where, for later use, we have introduced the notation $\mathcal{W}_{I_{1}}(I_1,J):=\partial\mathcal{W}(I_1,J)/\partial I_1$ and $\mathcal{W}_{J}(I_1,J):=\partial\mathcal{W}(I_1,J)/\partial J$.

For definiteness, unless otherwise stated, in all the simulations presented in this work, we make use of the values listed in Table \ref{Table1} for the material constants $\mu_r$, $\alpha_r$, and $\kappa$, which were deduced in \citep{Poulain17} to describe the elastic behavior of a popular silicone elastomer, PDMS Sylgard 184 with a 30:1 weight ratio of base elastomer to cross-linking agent. In the remainder of this paper, we simply refer to this synthetic elastomer as PDMS 30:1.

\subsubsection{The strength}\label{Sec: Strength}

Following \cite{KLP20}, the strength surface of the elastomer is taken to be characterized by the Drucker-Prager-type strength surface
\begin{equation}
\mathcal{F}(\bfS)=\sqrt{\dfrac{\mathcal{I}^2_1}{3}-\mathcal{I}_2}+\gamma_1 \mathcal{I}_1+\gamma_0=0\qquad {\rm with}\qquad \left\{\hspace{-0.1cm}\begin{array}{l}\gamma_0=-\dfrac{\sqrt{3}s_{\texttt{hs}} s_{\texttt{ts}}}
{3s_{\texttt{hs}}-s_{\texttt{ts}}}\vspace{0.2cm}\\
\gamma_1=\dfrac{s_{\texttt{ts}}}
{\sqrt{3}\left(3 s_{\texttt{hs}}-s_{\texttt{ts}}\right)}\end{array}\right. ,\label{DP}
\end{equation}
where
\begin{equation}
\left\{\begin{array}{l}
\mathcal{I}_1={\rm tr}\,\bfS^{(1)}=s_1+s_2+s_3\vspace{0.2cm}\\
\mathcal{I}_2=\dfrac{1}{2}\left[\left({\rm tr}\,\bfS^{(1)}\right)^2-{\rm tr}\,(\bfS^{(1)})^2\right]=\dfrac{1}{2}\left((s_1+s_2+s_3)^2-s_1^2-s_2^2-s_3^2\right)\end{array}\right.\label{SBiot-invariants}
\end{equation}
stand for the first and second principal invariants of the Biot stress tensor $\bfS^{(1)}=(\bfS^T\bfR+\bfR^T\bfS)/2$, while the material constants $s_{\texttt{ts}}>0$ and $s_{\texttt{hs}}>0$ denote the uniaxial tensile and hydrostatic strengths of the elastomer, that is, again, they denote the critical nominal stress values at which fracture nucleates under uniform states of monotonically increased uniaxial tension $\bfS={\rm diag}(s>0,0,0)$ and hydrostatic stress $\bfS={\rm diag}(s>0,s>0,s>0)$, respectively.

The two-material-parameter strength surface (\ref{DP}) is arguably the simplest model that allows to interpolate in a reasonable manner the few select strength data points that are typically available from experiments for elastomers, while at the same time it also permits to expediently explore different hydrostatic-to-uniaxial strength ratios $s_{\texttt{hs}}/s_{\texttt{ts}}$.

\begin{figure}[t!]
\centering
\includegraphics[width=0.80\linewidth]{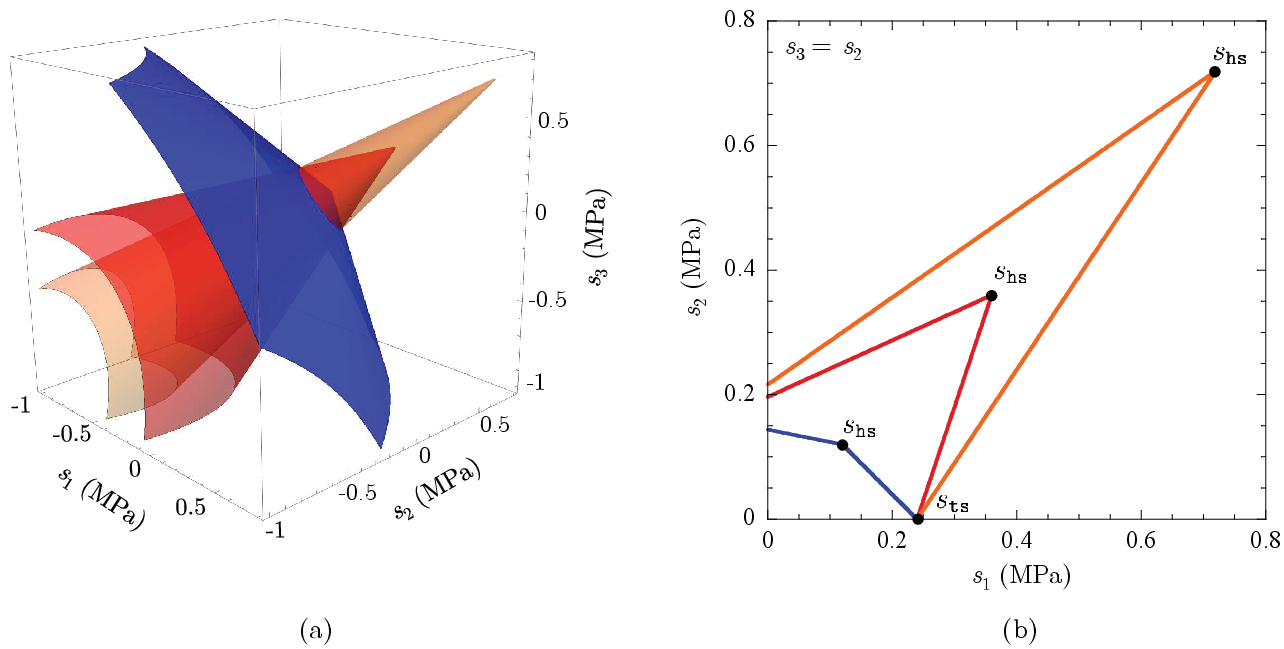}
\caption{{\small Strength surface (\ref{DP}) evaluated at the strength constants $s_{\texttt{ts}}$ and $s_{\texttt{hs}}$ listed in Table \ref{Table1}. (a) Plot of the surface in the space of principal Biot stresses $(s_1,s_2,s_3)$. (b) Plot of the surface in the space of principal stresses $(s_1,s_2)$ with $s_3=s_2$.}}\label{Fig2}
\end{figure}

Figure \ref{Fig2} shows plots, in the space of principal Biot stresses $(s_1,s_2,s_3)$, of the strength surface (\ref{DP}) evaluated at the values of $s_{\texttt{ts}}$ and $s_{\texttt{hs}}$ listed in Table \ref{Table1}. As determined in Appendix A of \citep{KLP20}, the values $s_{\texttt{ts}}=0.24$ MPa and $s_{\texttt{hs}}=0.36$ MPa are in the ballpark of the uniaxial tensile and hydrostatic strengths of PDMS 30:1. The two other values $s_{\texttt{hs}}=0.12$ MPa and $0.72$ MPa are selected here so as to lead to strength surfaces with a very small and a very large hydrostatic strength relative to the uniaxial tensile strength; recall that the case of a very weak hydrostatic strength relative to the uniaxial tensile strength is representative of natural rubber.

%
%

%
\begin{remark}
\emph{For given uniaxial tensile and hydrostatic strengths $\sts$ and $\shs$, the strength surface (\ref{DP}) predicts the biaxial tensile strength
\begin{equation*}
s_{\texttt{bs}}=\left(\dfrac{1}{3}+\dfrac{\shs}{\sts}\right)^{-1}\shs.
\end{equation*}
This relation implies in particular that $\shs>s_{\texttt{bs}}$ for $\shs/\sts>2/3$ and hence, for the values of $\sts$ and $\shs$ listed in Table \ref{Table1}, that $\shs>s_{\texttt{bs}}$ when $\shs>\sts$.
}
\end{remark}
\begin{remark}
\emph{The strength of any material, and hence of any elastomer, is inherently \emph{stochastic}. This is because the strength at a macroscopic material point $\bfX$ depends on the nature of the underlying defects from which fracture initiates, and this is known to exhibit a stochastic spatial variation in any given piece of material. Accordingly, the strength constants $\sts$ and $\shs$ in the strength surface (\ref{DP}) should be viewed not as deterministic but as stochastic material constants.}
\end{remark}
\begin{remark}
\emph{According to our choice of signs in (\ref{DP}), provided that $3 s_{\texttt{hs}}>s_{\texttt{ts}}$, any stress state such that
\begin{equation*}
\mathcal{F}(\bfS)\geq 0
\end{equation*}
is in violation of the strength of the elastomer.
}
\end{remark}

\subsubsection{The critical energy release rate}\label{Sec: Gc}

Finally, we take the critical energy release rate that describes the resistance to crack growth in the elastomer to be given by the \emph{constant}
\begin{equation*}
G_c.
\end{equation*}
For definiteness, in all the simulations presented in this work, we make use of the values listed in Table \ref{Table1} for $G_c$. Among these, the value $G_c=75$ N/m corresponds to a direct measurement by \cite{GT82} of a PDMS similar to PDMS 30:1. The additional values $G_c=10, 150, 200, 500$ N/m are selected here so as to probe the effect of the entire range of realistic values for the intrinsic fracture energy of common synthetic elastomers \citep{SLP23}.

\subsection{Boundary conditions}\label{Sec: BCs}

We consider that the specimen is firmly bonded to stiff fixtures which are then pulled apart quasistatically at a constant rate $\dot{h}_0$ so that the thickness of the specimen increases from $H$ to $h=H+\dot{h}_0 t$; see Fig. \ref{Fig1}.

Accordingly, we have the Dirichlet boundary condition
\begin{equation}
\bfy(\bfX,t)=\left\{\begin{array}{ll}
\overline{\bfy}_{\mathcal{T}}(\bfX,t)=X_1\bfe_1+X_2\bfe_2+\left(X_3+\dfrac{\dot{h}_0}{2}t\right)\bfe_3, &(\bfX,t)\in \partial\Omega_0^{\mathcal{T}}\times[0,T] \vspace{0.2cm}\\
\overline{\bfy}_{\mathcal{B}}(\bfX,t)=X_1\bfe_1+X_2\bfe_2+\left(X_3-\dfrac{\dot{h}_0}{2}t\right)\bfe_3, &(\bfX,t)\in \partial\Omega_0^{\mathcal{B}}\times[0,T] \end{array}\right.\label{BC-Dirichlet}
\end{equation}
at the top and bottom boundaries
\begin{equation*}
\partial\Omega_0^{\mathcal{T}}=\left\{\bfX:\sqrt{X_1^2+X_2^2}< \dfrac{D}{2},\, X_3=\dfrac{H}{2}\right\}\quad {\rm and}\quad \partial\Omega_0^{\mathcal{B}}=\left\{\bfX:\sqrt{X_1^2+X_2^2}< \dfrac{D}{2},\, X_3=-\dfrac{H}{2}\right\},
\end{equation*}
while the lateral boundary
\begin{equation*}
\partial\Omega_0^{\mathcal{L}}=\left\{\bfX:\sqrt{X_1^2+X_2^2}= \dfrac{D}{2}, \, |X_3|<\dfrac{H}{2}\right\}
\end{equation*}
is always traction free and hence subject to the Neumann boundary condition
\begin{equation}
\bfS(\bfX,t)\bfN=\textbf{0},\quad (\bfX,t)\in \partial\Omega_0^{\mathcal{L}}\times[0,T].\label{BC-Neumann}
\end{equation}

\subsection{The governing equations of elastostatics}

Prior to the nucleation of fracture, neglecting inertia and body forces, the combination of all of the above ingredients with the balance of linear momentum leads to the following set of governing equations of elastostatics

\begin{equation}
\left\{\begin{array}{ll}
{\rm Div}\left[\dfrac{\partial W}{\partial\bfF}(\nabla\bfy)\right]=\textbf{0},& \; (\bfX,t)\in\Omega_0\times[0,T]\vspace{0.2cm}\\
\bfy(\bfX,t)=\overline{\bfy}_{\mathcal{T}}(\bfX,t)=X_1\bfe_1+X_2\bfe_2+\left(X_3+\dfrac{\dot{h}_0}{2}t\right)\bfe_3, & \; (\bfX,t)\in\partial\Omega_0^{\mathcal{T}}\times[0,T]\vspace{0.2cm}\\
\bfy(\bfX,t)=\overline{\bfy}_{\mathcal{B}}(\bfX,t)=X_1\bfe_1+X_2\bfe_2+\left(X_3-\dfrac{\dot{h}_0}{2}t\right)\bfe_3, & \; (\bfX,t)\in\partial\Omega_0^{\mathcal{B}}\times[0,T]\vspace{0.2cm}\\
\left[\dfrac{\partial W}{\partial\bfF}(\nabla\bfy)\right]\bfN=\textbf{0},& \; (\bfX,t)\in\partial\Omega_0^{\mathcal{L}}\times[0,T]\vspace{0.2cm}\\
\bfy(\bfX,0)=\bfX,& \; \bfX\in\Omega_0
\end{array}\right.\label{Gov_Eqs}
\end{equation}
for the deformation field $\bfy(\bfX,t)$.

While the elastostatics equations (\ref{Gov_Eqs}) do not generally admit analytical solutions, they are amenable to numerical solution by the FE (finite element) method. So as to generate FE solutions that accurately and efficiently deal with the near incompressibility of the elastomer under study here --- note from Table \ref{Table1} that the ratio of initial bulk modulus $K=\kappa+2/3(\alpha_1\mu_1+\alpha_2\mu_2)$ to initial shear modulus $\mu=\mu_1+\mu_2$ is $K/\mu=10^3$ --- we make use of the non-conforming FE discretization of low order due to \cite{CR73}. All the FE results that are presented in this paper are generated with such a discretization.

\subsection{Where and when the strength surface is violated}

Having formulated the elastostatics problem (\ref{Gov_Eqs}), we are now in a position to investigate the pointwise stress field $\bfS(\bfX,t)$ in the poker-chip specimen during its loading, this prior to the nucleation of fracture. Because of the lack of loading singularities in the problem, such a stress field is warranted to be bounded almost everywhere, the sole exception being the geometric singularity at the curve
\begin{equation*}
\mathcal{C}_0=\left\{\bfX:\sqrt{X_1^2+X_2^2}=\dfrac{D}{2},\;|X_3|=\dfrac{H}{2}\right\},
\end{equation*}
where the free boundary of the specimen and the stiff fixtures meet. What is more, the spatial variation of the stress field turns out \emph{not} to be excessively large. This implies that the nucleation of fracture is expected to be dominated by the strength of the elastomer; see Remark \ref{Remark-bound} below.

Next, in order to begin gaining quantitative insight into the first instance of fracture nucleation in a poker-chip experiment, we present results for where and when the strength surface of the elastomer is violated in the specimen as the applied global deformation $\lambda=h/H$ increases, that is, for where and when
\begin{equation}
\mathcal{F}(\bfS)=\sqrt{\dfrac{\mathcal{I}^2_1}{3}-\mathcal{I}_2}+\dfrac{s_{\texttt{ts}}}
{\sqrt{3}\left(3 s_{\texttt{hs}}-s_{\texttt{ts}}\right)} \mathcal{I}_1-\dfrac{\sqrt{3}s_{\texttt{hs}} s_{\texttt{ts}}}
{3s_{\texttt{hs}}-s_{\texttt{ts}}}\geq 0.\label{DP-inequality}
\end{equation}
Aimed at separately identifying the effects of the properties of the elastomer and of the geometry of the poker-chip specimen, the results pertain to the three pairs of values $(s_{\texttt{ts}},s_{\texttt{hs}})=(0.24,0.12), (0.24,0.36), (0.24,0.72)$ MPa  for the uniaxial tensile and hydrostatic strengths listed in Table \ref{Table1}, which correspond to the ratios $s_{\texttt{hs}}/s_{\texttt{ts}}=1/2,3/2,3$, and the three different diameter-to-thickness ratios $D/H=40,10,4$. This particular set of representative values is selected here because it spans the entire range of possible scenarios.

\begin{remark}\label{Remark-bound}
\emph{Before proceeding with the presentation of the results, it is important to emphasize that satisfaction of the strength condition (\ref{DP-inequality}) at a material point $\bfX$ is a necessary but \emph{not} sufficient condition for fracture nucleation to occur at that point. This is because the stress field in a poker-chip experiment is \emph{non-uniform} and hence the nucleation of fracture in such an experiment is governed neither solely by strength nor solely by the Griffith competition between the elastic and fracture energies, but by an ``interpolation'' between the two; see Subsection 2.3 in \citep{KBFLP20}, \cite{LP23}, Appendix C in \citep{KLDLP24}, and Appendix B below. Nevertheless, because the non-uniformity of the stress field is not excessively large (in particular, save for the stress at the curve $\mathcal{C}_0$, the stress is not singular), the expectation is that the nucleation of fracture is dominated by the strength of the elastomer. This expectation was confirmed by \cite{KLP21} for natural rubber. The results in Sections \ref{Sec: 2D results} and \ref{Sec: Experiments} below show that this remains so for synthetic elastomers.
}
\end{remark}
%

%
\begin{figure}[t!]
\centering
\includegraphics[width=0.90\linewidth]{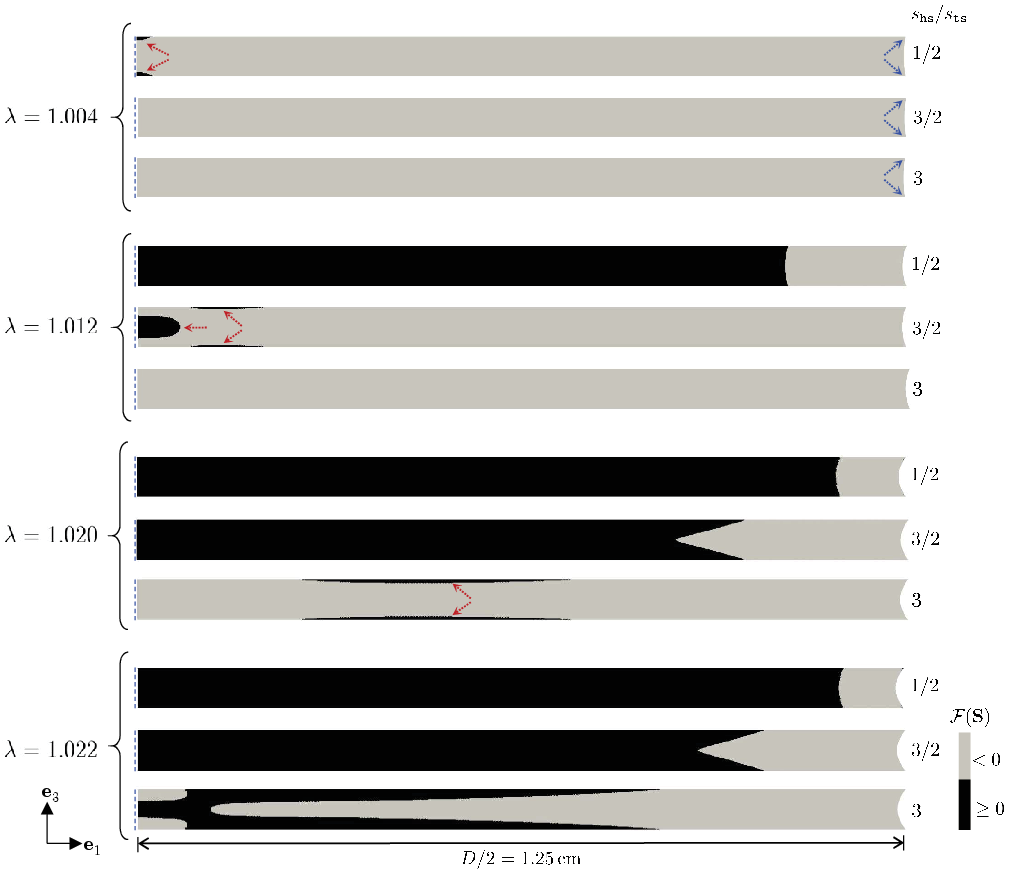}
\caption{{\small Contour plots, over the deformed configuration, of the regions of the specimen where the stress field exceeds ($\mathcal{F}(\bfS)\geq 0$) the Drucker-Prager-type strength surface of the elastomer at four increasing values of the normalized global deformation $\lambda=h/H$ between the fixtures. The results correspond to specimens with a diameter-to-thickness ratio $D/H=40$ and an elastomer with uniaxial tensile strength $s_{\texttt{ts}}=0.24$ MPa and the three different values of hydrostatic strength $s_{\texttt{hs}}$ listed in Table \ref{Table1}, so that $s_{\texttt{hs}}/s_{\texttt{ts}}=1/2, 3/2, 3$. The plots are shown only over half of the specimen for better visualization.}}\label{Fig3}
\end{figure}
%

%
\begin{figure}[t!]
\centering
\includegraphics[width=0.90\linewidth]{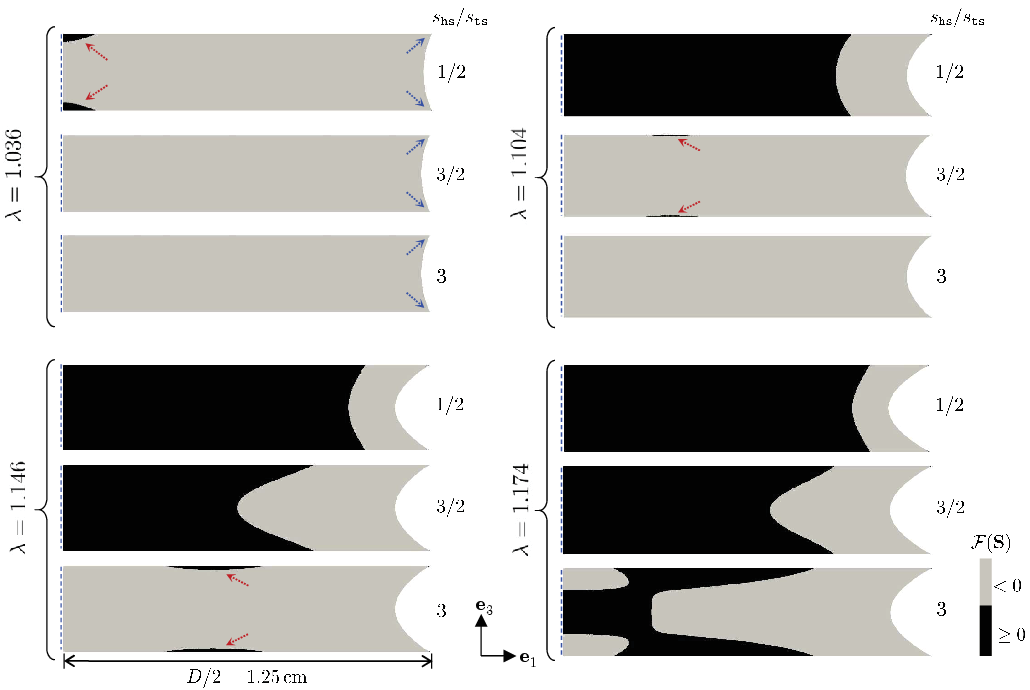}
\caption{{\small Contour plots, over the deformed configuration, of the regions of the specimen where the stress field exceeds ($\mathcal{F}(\bfS)\geq 0$) the Drucker-Prager-type strength surface of the elastomer at four increasing values of the normalized global deformation $\lambda=h/H$ between the fixtures. The results correspond to specimens with a diameter-to-thickness ratio $D/H=10$ and an elastomer with uniaxial tensile strength $s_{\texttt{ts}}=0.24$ MPa and the three different values of hydrostatic strength $s_{\texttt{hs}}$ listed in Table \ref{Table1}, so that $s_{\texttt{hs}}/s_{\texttt{ts}}=1/2, 3/2, 3$. The plots are shown only over half of the specimen for better visualization.}}\label{Fig4}
\end{figure}
%

%
\begin{figure}[t!]
\centering
\includegraphics[width=0.90\linewidth]{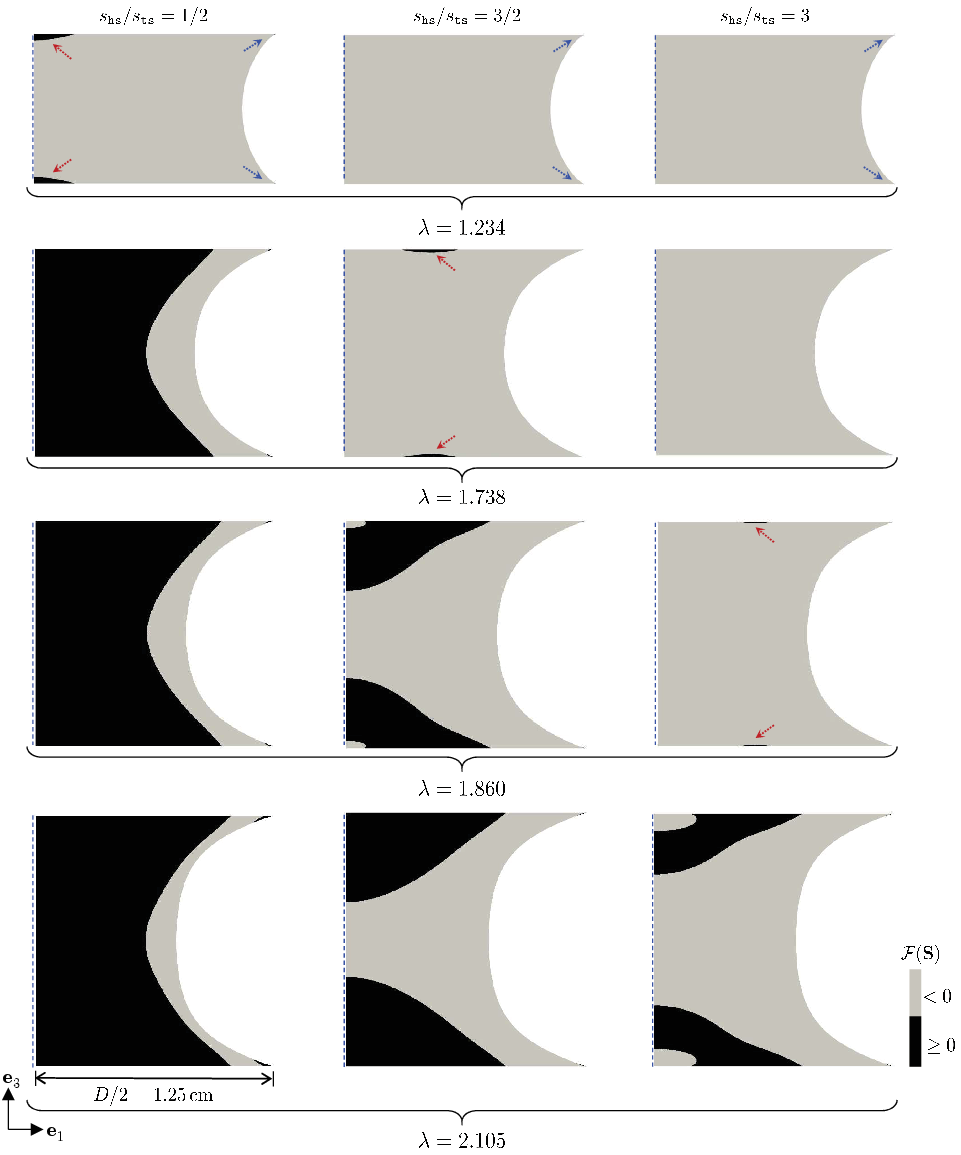}
\caption{{\small Contour plots, over the deformed configuration, of the regions of the specimen where the stress field exceeds ($\mathcal{F}(\bfS)\geq 0$) the Drucker-Prager-type strength surface of the elastomer at four increasing values of the normalized global deformation $\lambda=h/H$ between the fixtures. The results correspond to specimens with a diameter-to-thickness ratio $D/H=4$ and an elastomer with uniaxial tensile strength $s_{\texttt{ts}}=0.24$ MPa and the three different values of hydrostatic strength $s_{\texttt{hs}}$ listed in Table \ref{Table1}, so that $s_{\texttt{hs}}/s_{\texttt{ts}}=1/2, 3/2, 3$. The plots are shown only over half of the specimen for better visualization.}}\label{Fig5}
\end{figure}
%

Figure \ref{Fig3} presents contour plots of the regions (shown in black) of the specimen at which the criterion (\ref{DP-inequality}) is satisfied, and hence at which the strength of the elastomer is exceeded, at four increasing values $\lambda$ of the normalized global deformation between the fixtures. The results are shown over the deformed configuration and pertain to a very thin specimen with initial thickness $H=0.0625$ cm and hence diameter-to-thickness ratio $D/H=40$. As for all the simulations presented in the main body of this work, again, the elasticity of the elastomer is described by the stored-energy function (\ref{W-LP}) with the material constants listed in Table \ref{Table1}. For direct comparison, the figure shows plots for all the three pairs of values $(s_{\texttt{ts}},s_{\texttt{hs}})=(0.24,0.12), (0.24,0.36), (0.24,0.72)$ MPa  for the uniaxial tensile and hydrostatic strengths listed in Table \ref{Table1}.

There are several key observations worth remarking from the results presented in Fig. \ref{Fig3}. First, irrespective of the value of the hydrostatic-to-uniaxial strength ratio $s_{\texttt{hs}}/s_{\texttt{ts}}$, the first violation of the strength surface always happens at the curve $\mathcal{C}_0$, where, again, the free boundary of the specimen and the stiff fixtures meet (indicated by the blue arrows). Such a strength violation happens instantly, at the first increment in the applied deformation $\lambda$. This is because the curve $\mathcal{C}_0$ is a geometric singularity. Importantly, the region in the specimen where the strength surface is initially violated is confined to within a very small neighborhood around $\mathcal{C}_0$ and remains so for increasing values of the applied deformation.

Second, away from $\mathcal{C}_0$, the violation of the strength surface occurs first at the boundary between the elastomer and the fixtures (indicated by the red arrows). Crucially, for the elastomer with the smallest hydrostatic-to-uniaxial strength ratio $s_{\texttt{hs}}/s_{\texttt{ts}}=1/2$, this initial region of strength violation is perfectly aligned with the centerline of the specimen. For the elastomers with the larger hydrostatic-to-uniaxial strength ratios $s_{\texttt{hs}}/s_{\texttt{ts}}=3/2$ and $3$, on the other hand, the regions where the strength surface is first violated are \emph{radially away from the centerline}, roughly, at distances $D/20$ and $D/5$ away, respectively. Interestingly, for the elastomer with the intermediate strength ratio $s_{\texttt{hs}}/s_{\texttt{ts}}=3/2$, the strength surface is also first violated at the center of the specimen (also indicated by a red arrow). Upon further loading, the regions at the boundary between the elastomer and the fixtures where the strength surface is first violated grow in size both radially and towards the midplane of the specimen.

Third, the state of stress in all the regions in the specimen where the strength surface is violated (i.e., all the black regions) is \emph{triaxial} and \emph{all tensile} $(s_1>0,s_2>0,s_3>0)$ with $s_1\neq s_2\neq s_3\neq s_1$. In other words, the violation of the strength surface $\mathcal{F}(\bfS)=0$ always takes place in the first octant in the space of principal stresses $(s_1,s_2,s_3)$.

Finally, again away from $\mathcal{C}_0$, the critical values of the applied global deformation $\lambda$ at which the strength surface is first violated are roughly $\lambda=1.004, 1.012, 1.020$ for the elastomers with the strength ratios $s_{\texttt{hs}}/s_{\texttt{ts}}=1/2, 3/2, 3$, respectively. That is, significantly larger deformations $\lambda$ are required to exceed the strength of elastomers with the same uniaxial tensile strength but larger hydrostatic strengths.

Figures \ref{Fig4} and \ref{Fig5} present contour plots analogous to those of Fig. \ref{Fig3} for thicker specimens with initial thicknesses $H=0.25$ cm and $0.625$ cm and hence diameter-to-thickness ratios $D/H=10$ and $4$. Qualitatively, where and how the strength surface is first violated and proceeds to be violated throughout the specimen upon further loading is similar to that shown in Fig. \ref{Fig3} for the very thin specimen. Accordingly, away from the singular curve $\mathcal{C}_0$, the violation of the strength surface occurs first at the boundary between the elastomer and the fixtures (indicated by the red arrows). For the elastomer with the smallest strength ratio $s_{\texttt{hs}}/s_{\texttt{ts}}=1/2$, this initial region of strength violation is aligned with centerline of the specimen. For the elastomers with the larger strength ratios $s_{\texttt{hs}}/s_{\texttt{ts}}=3/2$ and $3$, the initial regions of strength violation are radially away from the centerline. Moreover, for a given diameter-to-thickness ratio $D/H$, significantly larger deformations are requited to exceed the strength of elastomers with larger hydrostatic strengths.

In summary, the above elementary strength analysis suggests that the first nucleation of fracture in a poker-chip experiment can occur either along the centerline of the specimen or radially away from it. Specifically, nucleation of fracture along the centerline is more likely to occur in elastomers with a smaller hydrostatic-to-uniaxial strength ratio $s_{\texttt{hs}}/s_{\texttt{ts}}$ (such as natural rubber), while nucleation of fracture radially away from the centerline is more likely to occur in elastomers with larger $s_{\texttt{hs}}/s_{\texttt{ts}}$ (such as PDMS 30:1). What is more, the analysis has revealed that, irrespective of its location, nucleation of fracture will always occur in a region where the strength of the elastomer is violated in a state of \emph{all-tensile triaxial stress} $(s_1>0,s_2>0,s_3>0)$ with $s_1\neq s_2\neq s_3\neq s_1$, and hence \emph{not} one of identically hydrostatic stress $(s>0,s>0,s>0)$.

Now, as emphasized in Remark \ref{Remark-bound} above, the violation (\ref{DP-inequality}) of the strength surface of the elastomer, which is the focus of the above analysis, is a necessary but \emph{not} sufficient condition for the nucleation of fracture. To determine where and when fracture actually nucleates and propagates in a poker-chip experiment one needs to make use of a complete theory of fracture. We do just that in the next sections.

\section{A complete fracture nucleation and propagation analysis of the poker-chip experiment}\label{Sec: The theory}

In addition to the deformation field $\bfy(\bfX,t)$, the boundary conditions (\ref{BC-Dirichlet})-(\ref{BC-Neumann}) applied in a poker-chip experiment eventually result in the nucleation and subsequent propagation of cracks in the specimen. We describe such cracks in a regularized fashion via an order parameter or phase field
\begin{equation*}
z=z(\bfX, t)
\end{equation*}
taking values in the range $[0,1]$. The value $z=1$ identifies the intact regions of the elastomer and $z=0$ those that have been fractured, while the transition from $z=1$ to $z=0$ is set to occur smoothly over regions of small thickness of regularization length scale $\varepsilon>0$.

\subsection{The governing equations of deformation and fracture in their general form}

According to the phase-field fracture formulation put forth by \citet*{KFLP18}, the deformation field $\bfy_k(\bfX)=\bfy(\bfX,t_k)$ and phase field $z_k(\bfX)=z(\bfX,t_k)$ at any material point $\bfX \in \overline{\Omega}_0=\Omega_0\cup\partial\Omega_0$ and at any given discrete time $t_k\in\{0=t_0,t_1,...,t_m,t_{m+1},...,t_M=T\}$ are determined by the system of coupled partial differential equations (PDEs)

\begin{equation}
\left\{\begin{array}{ll}
{\rm Div}\left[z_{k}^2\dfrac{\partial W}{\partial \bfF}(\nabla \bfy_{k})\right]=\textbf{0},& \,\bfX\in\Omega_0\vspace{0.2cm}\\
\bfy_k(\bfX)=\overline{\bfy}_{\mathcal{T}}(\bfX,t_k), & \; \bfX\in\partial\Omega_0^{\mathcal{T}}\vspace{0.2cm}\\
\bfy_k(\bfX)=\overline{\bfy}_{\mathcal{B}}(\bfX,t_k), & \; \bfX,\in\partial\Omega_0^{\mathcal{B}}\vspace{0.2cm}\\
\left[z_k^2\dfrac{\partial W}{\partial\bfF}(\nabla\bfy_k)\right]\bfN=\textbf{0},& \; \bfX\in\partial\Omega_0^{\mathcal{L}}
\end{array}\right. \label{BVP-y-theory}
\end{equation}
and
\begin{equation}
\left\{\begin{array}{ll}
\hspace{-0.15cm} {\rm Div}\left[\varepsilon\, G_c \nabla z_k\right]=\dfrac{8}{3}z_{k} W(\nabla\bfy_k)-\dfrac{4}{3}c_\texttt{e}(\bfX,t_{k})-\dfrac{G_c}{2\varepsilon},&\, \mbox{ if } z_{k}(\bfX)< z_{k-1}(\bfX),\quad \bfX\in \Omega_0
\vspace{0.2cm}\\
\hspace{-0.15cm}
{\rm Div}\left[\varepsilon\, G_c \nabla z_k\right]\geq\dfrac{8}{3}z_{k} W(\nabla\bfy_k)-\dfrac{4}{3}c_\texttt{e}(\bfX,t_{k})-\dfrac{G_c}{2\varepsilon},&\, \mbox{ if } z_{k}(\bfX)=1\; \mbox{ or }\; z_{k}(\bfX)= z_{k-1}(\bfX)>0, \quad \bfX\in \Omega_0\vspace{0.2cm}\\
\hspace{-0.15cm}
z_k(\bfX)=0,&\, \mbox{ if } z_{k-1}(\bfX)=0, \quad \bfX\in \Omega_0\vspace{0.2cm}\\
\hspace{-0.15cm}\nabla z_k\cdot\bfN=0,& \, \bfX\in \partial\Omega_0
\end{array}\right. \label{BVP-z-theory}
\end{equation}
with $\bfy(\bfX,0)\equiv\bfX$ and $z(\bfX,0)\equiv1$. In these equations, we recall that the stored-energy function $W(\bfF)$ is given by expression (\ref{W-LP}), the  deformations $\overline{\bfy}_{\mathcal{T}}(\bfX,t)$ and $\overline{\bfy}_{\mathcal{B}}(\bfX,t)$ applied at the top and bottom boundaries of the specimen are given by expressions (\ref{BC-Dirichlet}), $\nabla z_k=\nabla z(\bfX,t_k)$, and $c_\texttt{e}(\bfX,t)$ is a driving force whose specific constitutive prescription, as elaborated next, depends on the particular form of strength surface $\mathcal{F}(\bfS)=0$.

\begin{remark}\label{Remark: irreversibility}
{\rm The inequalities in (\ref{BVP-z-theory}) stem from the facts that, by definition, the phase field is bounded according to $0\leq z\leq 1$ and, by constitutive assumption, fracture is an irreversible process, in other words, healing is not allowed. Recent experimental evidence has revealed that internally nucleated cracks in some elastomers may self-heal \citep{Poulain17,Poulain18}. The inequalities (\ref{BVP-z-theory}) can be augmented to describe such a healing process, but we shall not consider healing in this work; see Subsection 3.2 in \citep{KFLP18} for the relevant details and \citep{FGLP19} for the corresponding ``sharp-theory'' perspective.
}
\end{remark}

\begin{remark}{\rm The parameter $\varepsilon$ in (\ref{BVP-z-theory}), with units of $length$, regularizes sharp cracks. Accordingly, by definition, it can be arbitrarily small. In practice, $\varepsilon$ should be selected to be smaller than the smallest material length scale built in (\ref{BVP-y-theory})-(\ref{BVP-z-theory}),  which comes about because of the different units of the elastic stored-energy function $W(\bfF)$ ($force/length^2$), the strength function $\mathcal{F}(\bfS)$ ($force/length^2$), and the critical energy release rate $G_c$ ($force/length$); see Appendix B below. As a rule of thumb, it typically suffices to set $\varepsilon<3G_c/(16\mathcal{W}_{\texttt{ts}})$, where $\mathcal{W}_{\texttt{ts}}$ is given by expression (\ref{WtsWhs})$_1$ below.
}\label{Remark_length}
\end{remark}

\subsection{A new constitutive prescription for the driving force $c_\texttt{\emph{e}}(\bfX,t)$}\label{Sec: New ce}

Given a strength surface $\mathcal{F}(\bfS)=0$, \cite{KBFLP20} and \cite{KLP20} provided a blueprint to construct constitutive prescriptions for the driving force $c_\texttt{e}(\bfX,t)$ and made use of that blueprint to work out a particular constitutive prescription for the case when the strength surface is of Drucker-Prager type.
In this subsection, we make use of the same blueprint to work out a new constitutive prescription for $c_\texttt{e}(\bfX,t)$ that has additional advantages over the original one, chief among them that it is fully explicit.

\subsubsection{The general functional form of $c_{\texttt{\emph{e}}}(\bfX,t)$}

As in the preceding section, we assume that the strength surface of the elastomer is characterized by the Drucker-Prager-type strength surface (\ref{DP}). Accordingly, following \cite{KBFLP20}, \cite{KLP20}, and \cite{KRLP22}, we begin by considering driving forces of the Drucker-Prager form
\begin{align}
c_{\texttt{e}}(\bfX,t)=\beta_2^\varepsilon\sqrt{\dfrac{\mathcal{I}^2_1}{3}-\mathcal{I}_2}+\beta_1^\varepsilon  \mathcal{I}_1+\beta_0^\varepsilon+z\left(1-\dfrac{\sqrt{\mathcal{I}^2_1}}{\mathcal{I}_1}\right)\mathcal{W}(I_1,J)+\dfrac{3}{4}{\rm Div}\left[\varepsilon(\delta^{\varepsilon}-1) G_c\nabla z \right],\label{cehat}
\end{align}
where $\beta_0^\varepsilon$, $\beta_1^\varepsilon$, $\beta_2^\varepsilon$, and $\delta^{\varepsilon}$,  are $\varepsilon$-dependent coefficients, to be spelled out below, and  $\mathcal{I}_1$ and $\mathcal{I}_2$ stand for the first two principal invariants (\ref{SBiot-invariants}) of the Biot stress tensor
\begin{align*}
\bfS^{(1)}(\bfX,t)=z^2\dfrac{\partial \mathcal{W}}{\partial\bfU}(I_1,J)=&z^2\left[2\mathcal{W}_{I_{1}}\bfU+\mathcal{W}_{J}J\bfU^{-1}\right]\nonumber\\
=&z^2\left[\left(\sum_{r=1}^{2}3^{1-\alpha_{r}}\mu_r I_1^{\alpha_r-1}\right)\bfU+\left(\kappa(J-1)-\sum_{r=1}^{2}\dfrac{\mu_r}{J}\right)J\bfU^{-1}\right]
\end{align*}
and hence read as
\begin{equation}
\left\{\begin{array}{l}
\mathcal{I}_1=2 z^2 i_1 \mathcal{W}_{I_1}+z^2 i_2 \mathcal{W}_{J}\vspace{0.1cm}\\
\hspace{0.45cm} =z^2 i_1 \left(\sum_{r=1}^{2}3^{1-\alpha_{r}}\mu_r I_1^{\alpha_r-1}\right)+z^2 i_2 \left(\kappa(J-1)-\sum_{r=1}^{2}\dfrac{\mu_r}{J}\right) \vspace{0.3cm}\\
\mathcal{I}_2=4 z^4 i_2 \mathcal{W}^2_{I_1}+ z^4 i_1 J \mathcal{W}^2_{J}+2 z^4 (i_1 i_2-3 J) \mathcal{W}_{J}\mathcal{W}_{I_1}\vspace{0.1cm}\\
\hspace{0.45cm} =z^4 i_2 \left(\sum_{r=1}^{2}3^{1-\alpha_{r}}\mu_r I_1^{\alpha_r-1}\right)^2+ z^4 i_1 J \left(\kappa(J-1)-\sum_{r=1}^{2}\dfrac{\mu_r}{J}\right)^2+\vspace{0.1cm}\\
\hspace{0.75cm} z^4 (i_1 i_2-3 J) \left(\kappa(J-1)-\sum_{r=1}^{2}\dfrac{\mu_r}{J}\right)\left(\sum_{r=1}^{2}3^{1-\alpha_{r}}\mu_r I_1^{\alpha_r-1}\right) \end{array}\right. ,\label{I1I2-Biot}
\end{equation}
where $i_1$ and $i_2$ stand for the first two principal invariants of the right stretch tensor $\bfU$:
\begin{equation*}
i_1={\rm tr}\,\bfU\quad{\rm and}\quad i_2=\dfrac{1}{2}\left[({\rm tr}\,\bfU)^2-{\rm tr}\,\bfU^2\right].
\end{equation*}
\begin{remark}
\emph{The stress invariants (\ref{I1I2-Biot}) depend on the phase field $z(\bfX,t)$ directly and on the deformation field $\bfy(\bfX,t)$ via the invariants $I_1$ and $J$ of the right Cauchy-Green deformation tensor $\bfC(\bfX,t)=\nabla\bfy^T(\bfX,t)\nabla\bfy(\bfX,t)$ and the invariants $i_1$ and $i_2$ of the right stretch tensor $\bfU(\bfX,t)=\sqrt{\nabla\bfy^T(\bfX,t)\nabla\bfy(\bfX,t)}$. To circumvent having to perform numerically the polar decomposition of the deformation gradient $\nabla\bfy(\bfX,t)$ to compute $i_1$ and $i_2$, we recall here that $i_1$ and $i_2$ can be written explicitly in terms of $I_1$, $J$, and $I_2=\left(({\rm tr}\bfC)^2-{\rm tr}\bfC^2\right)/2$. The pertinent relations read
\begin{align*}
\left\{\begin{array}{l}
i_1= \left\{\begin{array}{ll}\dfrac{1}{2}\left(\sqrt{2 I_1+\xi_3}+\sqrt{2 I_1-\xi_3+\dfrac{16 J}{\sqrt{2 I_1+\xi_3}}}\right)& {\rm if}\;\xi_3\neq -2 I_1 \\ \\
\sqrt{I_1+2\sqrt{I_2}}& {\rm if}\;\xi_3 = -2 I_1 \end{array}\right.\\ \\
i_2= \sqrt{I_2+2 i_1 J} \end{array}\right. \label{i-I},
\end{align*}
where
\begin{equation*}
\xi_3=-\dfrac{2}{3}I_1+\left(\xi_1+\sqrt{\xi_2}\right)^{1/3}+\left(\xi_1-\sqrt{\xi_2}\right)^{1/3}
\end{equation*}
with
\begin{equation*}
\xi_1=\dfrac{2^5}{27}\left(2 I_1^3-9 I_1 I_2+27 J^2\right)\quad {\rm and}\quad \xi_2=\dfrac{2^{10}}{27}\left(4 I_2^3-I_1^2 I_2^2+4 I_1^3 J^2-18 I_1 I_2 J^2+27 J^4\right);
\end{equation*}
see, e.g., \cite{Carlson84} and \cite{Steigmann02}.
}
\end{remark}

\subsubsection{The asymptotic form of the coefficients $\beta_0^\varepsilon$, $\beta_1^\varepsilon$, $\beta_2^\varepsilon$ in the limit as $\varepsilon\searrow 0$}

The coefficients $\beta_0^\varepsilon$, $\beta_1^\varepsilon$, $\beta_2^\varepsilon$ in the driving force (\ref{cehat}) must be selected so that for \emph{uniform} deformation gradients $\nabla\bfy(\bfX,t)=\overline{\bfF}$ and \emph{uniform} phase field $z(\bfX,t)=1$ throughout the entire domain $\Omega_0$ occupied by the specimen, the evolution equation (\ref{BVP-z-theory})$_1$ for the phase field is satisfied precisely when the given strength surface $\mathcal{F}(\bfS)=0$, with $\bfS=\partial W(\overline{\bfF})/\partial\bfF$, is first violated in the limit as the regularization length $\varepsilon\searrow 0$. Precisely, we require that
\begin{align}
-\dfrac{4}{3}\left(\beta_2^\varepsilon\sqrt{\dfrac{\mathcal{I}^2_1}{3}-\mathcal{I}_2}+\beta_1^\varepsilon  \mathcal{I}_1+\beta_0^\varepsilon\right)-\dfrac{G_c}{2\varepsilon}=&C\mathcal{F}(\bfS)\nonumber\\
=&C\left(\sqrt{\dfrac{\mathcal{I}^2_1}{3}-\mathcal{I}_2}+\dfrac{s_{\texttt{ts}}}
{\sqrt{3}\left(3 s_{\texttt{hs}}-s_{\texttt{ts}}\right)} \mathcal{I}_1-\dfrac{\sqrt{3}s_{\texttt{hs}} s_{\texttt{ts}}}
{3s_{\texttt{hs}}-s_{\texttt{ts}}}\right)\label{Cond-F-0}
\end{align}
for some constant $C$ in the limit as $\varepsilon\searrow 0$. Clearly, the choice
\begin{equation}
\left\{\begin{array}{l}\beta_0^\varepsilon=(\delta^\varepsilon-1)\dfrac{3 G_c}{8 \varepsilon}\vspace{0.2cm}\\
\beta^\varepsilon_1=-\dfrac{1}{\shs}\delta^\varepsilon\dfrac{G_c}{8\varepsilon}\vspace{0.2cm}\\
\beta^\varepsilon_2=-\dfrac{\sqrt{3}(3\shs-\sts)}{\shs\sts}\delta^\varepsilon\dfrac{G_c}{8\varepsilon}\end{array}\right. , \label{betas-0}
\end{equation}
where $\delta^\varepsilon$ is, for now, an arbitrary coefficient, satisfies condition (\ref{Cond-F-0}).

\subsubsection{The complete expressions for the coefficients $\beta_0^\varepsilon$, $\beta_1^\varepsilon$, $\beta_2^\varepsilon$}

However small, the value of the regularization length $\varepsilon$ is always finite in practice. It is thus desirable to choose coefficients in the constitutive prescription (\ref{cehat}) for the driving force that result in a phase-field theory that is capable of describing accurately the strength surface of the elastomer  not only asymptotically in the limit as $\varepsilon\searrow 0$ but also at finite values of $\varepsilon$. This can be accomplished by adding corrections of $O(\varepsilon^0)$ and higher orders to the expressions for $\beta_1^\varepsilon$ and $\beta_2^\varepsilon$ in (\ref{betas-0}).

Here, to determine the corrections in $\beta_1^\varepsilon$ and $\beta_2^\varepsilon$, we require that when the entire domain $\Omega_0$ occupied by the specimen is subjected to \emph{uniform} uniaxial tension $\bfS={\rm diag}(s>0,0,0)$ and \emph{uniform} hydrostatic loading $\bfS={\rm diag}(s>0,s>0,s>0)$, when the phase field is also \emph{uniformly} $z(\bfX,t)=1$, the evolution equation (\ref{BVP-z-theory})$_1$ is satisfied precisely when the given strength surface $\mathcal{F}(\bfS)=0$ is violated, this for any value of $\varepsilon$. Precisely, we require that
\begin{equation}
\left\{\begin{array}{l}
\dfrac{8}{3}\mathcal{W}_{\texttt{ts}}-\dfrac{4}{3}\left(\beta_2^\varepsilon\dfrac{\sts}{\sqrt{3}}+\beta_1^\varepsilon  \sts+\beta_0^\varepsilon\right)-\dfrac{G_c}{2\varepsilon}=0\vspace{0.2cm}\\
\dfrac{8}{3}\mathcal{W}_{\texttt{hs}}-\dfrac{4}{3}\left(3\beta_1^\varepsilon  \shs+\beta_0^\varepsilon\right)-\dfrac{G_c}{2\varepsilon}=0\end{array}\right.,\label{Cond-F-1}
\end{equation}
where $\mathcal{W}_{\texttt{ts}}$ and $\mathcal{W}_{\texttt{hs}}$ stand for the values of the stored-energy function (\ref{W-LP}) along uniform uniaxial tension and hydrostatic stress states at which the strength surface is violated. To wit,
\begin{equation}
\mathcal{W}_{\texttt{ts}}=\mathcal{W}(\l^2_{\texttt{ts}}+2\l^2_l, \l_{\texttt{ts}}\l_l^2)\qquad {\rm and}\qquad \mathcal{W}_{\texttt{hs}}=\mathcal{W}(3\l^2_{\texttt{hs}}, \l^3_{\texttt{hs}}),\label{WtsWhs}
\end{equation}
where the pair of stretches ($\l_{\texttt{ts}}, \l_l$) and the stretch $\l_{\texttt{hs}}$ are defined implicitly as the roots closest to $(1,1)$ and $1$ of the system of nonlinear algebraic equations
\begin{align*}
&\left\{\begin{array}{l}
s_{\texttt{ts}}=2 \l_{\texttt{ts}}\mathcal{W}_{I_1}(\l^2_{\texttt{ts}}+2\l^2_l, \l_{\texttt{ts}}\l_l^2)+\l^2_l\mathcal{W}_{J}(\l^2_{\texttt{ts}}+2\l^2_l, \l_{\texttt{ts}}\l_l^2)\\[10pt]
0=2\l_l\mathcal{W}_{I_1}(\l^2_{\texttt{ts}}+2\l^2_l, \l_{\texttt{ts}}\l_l^2)+\l_{\texttt{ts}}\l_l\mathcal{W}_{J}(\l^2_{\texttt{ts}}+2\l^2_l, \l_{\texttt{ts}}\l_l^2)\end{array}\right.  
\end{align*}
and the nonlinear algebraic equation
\begin{equation*}
s_{\texttt{hs}}=2\l_{\texttt{hs}}\mathcal{W}_{I_1}(3\l^2_{\texttt{hs}}, \l^3_{\texttt{hs}})+\l^2_{\texttt{hs}}\mathcal{W}_{J}(3\l^2_{\texttt{hs}}, \l^3_{\texttt{hs}}),
\end{equation*}
respectively.

For the stored-energy function (\ref{W-LP}) with the material constants listed in Table \ref{Table1}, the values of $\mathcal{W}_{\texttt{ts}}$  and $\mathcal{W}_{\texttt{hs}}$ read
\begin{equation*}
 \mathcal{W}_{\texttt{ts}}=0.341\; {\rm MPa} \qquad {\rm and}\qquad \mathcal{W}_{\texttt{hs}}=\left\{\begin{array}{ll}
0.14\times 10^{-3}\; {\rm MPa} & {\rm if}\,\; \shs=0.12 \; {\rm MPa}\vspace{0.2cm}\\
1.27\times 10^{-3}\; {\rm MPa} & {\rm if}\,\; \shs=0.36 \; {\rm MPa}\vspace{0.2cm}\\
5.04\times 10^{-3}\; {\rm MPa} &{\rm if}\,\; \shs=0.72 \; {\rm MPa}\\ \end{array}\right..
\end{equation*}
Note that the very small values of $\mathcal{W}_{\texttt{hs}}$ are nothing more than the manifestation of the fact that the silicone elastomer (PDMS 30:1) described by the stored-energy function (\ref{W-LP}) with the material constants listed in Table \ref{Table1} is nearly incompressible. For nearly incompressible elastomers, one can safely use the approximation $\mathcal{W}_{\texttt{hs}}=0$ at little loss of accuracy.

At this stage, it is straightforward to deduce that the choice
\begin{equation}
\left\{\begin{array}{l}\beta_0^\varepsilon=(\delta^\varepsilon-1)\dfrac{3 G_c}{8 \varepsilon}\vspace{0.2cm}\\
\beta^\varepsilon_1=-\dfrac{1}{\shs}\delta^\varepsilon\dfrac{G_c}{8\varepsilon}+\dfrac{2\mathcal{W}_{\texttt{hs}}}{3\shs}\vspace{0.2cm}\\
\beta^\varepsilon_2=-\dfrac{\sqrt{3}(3\shs-\sts)}{\shs\sts}\delta^\varepsilon\dfrac{G_c}{8\varepsilon}-
\dfrac{2\mathcal{W}_{\texttt{hs}}}{\sqrt{3}\shs}+\dfrac{2\sqrt{3}\mathcal{W}_{\texttt{ts}}}{\sts}\end{array}\right. , \label{betas}
\end{equation}
where $\delta^\varepsilon$ is still an arbitrary coefficient, satisfies the asymptotic condition (\ref{Cond-F-0}) in the limit as $\varepsilon\searrow 0$, as well as the conditions (\ref{Cond-F-1}) for any value of $\varepsilon$. Expressions (\ref{betas}) are hence the coefficients that we are after for the constitutive prescription (\ref{cehat}).

\subsubsection{The resulting expression for $c_{\texttt{\emph{e}}}(\bfX,t)$}

Substitution of the coefficients (\ref{betas}) in expression (\ref{cehat}) leads to the external driving force
\begin{align}
c_{\texttt{e}}(\bfX,t)=\widehat{c}_{\texttt{e}}(\bfX,t)+(\delta^\varepsilon-1)\dfrac{3 G_c}{8 \varepsilon}+\dfrac{3}{4}{\rm Div}\left[\varepsilon (\delta^{\varepsilon}-1) G_c\nabla z \right]\label{ce-Final}
\end{align}
with
\begin{align}
\widehat{c}_{\texttt{e}}(\bfX,t)=&\left(-\dfrac{\sqrt{3}(3\shs-\sts)}{\shs\sts}\delta^\varepsilon\dfrac{G_c}{8\varepsilon}-
\dfrac{2\mathcal{W}_{\texttt{hs}}}{\sqrt{3}\shs}+\dfrac{2\sqrt{3}\mathcal{W}_{\texttt{ts}}}{\sts}\right)\sqrt{\dfrac{\mathcal{I}^2_1}{3}-\mathcal{I}_2}+\left(-\dfrac{1}{\shs}\delta^\varepsilon\dfrac{G_c}{8\varepsilon}+\dfrac{2\mathcal{W}_{\texttt{hs}}}{3\shs}\right)  \mathcal{I}_1+\nonumber\\
&z\left(1-\dfrac{\sqrt{\mathcal{I}^2_1}}{\mathcal{I}_1}\right)\mathcal{W}(I_1,J),\label{cehat-Final}
\end{align}
where we recall that $\mathcal{I}_1$ and $\mathcal{I}_2$ are given by the explicit expressions (\ref{I1I2-Biot}) in terms of the phase field $z(\bfX,t)$ and the deformation field $\bfy(\bfX,t)$.

Note that the constitutive prescription (\ref{ce-Final}) with (\ref{cehat-Final}) depends directly on the stored-energy function $W(\bfF)$ via $\mathcal{I}_1$, $\mathcal{I}_2$, $\mathcal{W}_{\texttt{ts}}$, $\mathcal{W}_{\texttt{hs}}$, the strength material constants $\sts$, $\shs$, and the critical energy release rate $G_c$ of the elastomer of interest, as well as on the regularization length $\varepsilon$.

Note also that (\ref{ce-Final})-(\ref{cehat-Final}) is a fully explicit expression \emph{up to} the prescription of the coefficient $\delta^\varepsilon$. The prescription of this coefficient is detailed in Subsection \ref{Sec: delta} below.

\subsection{The governing equations}

Making use of the constitutive prescription (\ref{ce-Final}) with (\ref{cehat-Final}) for the driving force $c_{\texttt{e}}(\bfX,t)$ in the general form (\ref{BVP-y-theory})-(\ref{BVP-z-theory}) of the governing equations, as well as of the penalty method to regularize the inequalities in (\ref{BVP-z-theory}) enforcing that the phase field remains in the physically admissible range $0\leq z\leq 1$ and that fracture is irreversible, leads to the final form of the governing equations

\begin{equation}
\left\{\begin{array}{ll}
{\rm Div}\left[z_{k}^2\dfrac{\partial W}{\partial \bfF}(\nabla \bfy_{k})\right]=\textbf{0},& \,\bfX\in\Omega_0\vspace{0.2cm}\\
\bfy_k(\bfX)=\overline{\bfy}_{\mathcal{T}}(\bfX,t_k), & \; \bfX\in\partial\Omega_0^{\mathcal{T}}\vspace{0.2cm}\\
\bfy_k(\bfX)=\overline{\bfy}_{\mathcal{B}}(\bfX,t_k), & \; \bfX,\in\partial\Omega_0^{\mathcal{B}}\vspace{0.2cm}\\
\left[z_k^2\dfrac{\partial W}{\partial\bfF}(\nabla\bfy_k)\right]\bfN=\textbf{0},& \; \bfX\in\partial\Omega_0^{\mathcal{L}}
\end{array}\right. \label{BVP-y-theory-reg}
\end{equation}
and
\begin{equation}
\left\{\begin{array}{ll}
\hspace{-0.15cm} {\rm Div}\left[\varepsilon\, \delta^\varepsilon G_c \nabla z_k\right]=\dfrac{8}{3}z_{k} W(\nabla\bfy_k)-\dfrac{4}{3}\widehat{c}_\texttt{e}(\bfX,t_{k})-\dfrac{\delta^\varepsilon G_c}{2\varepsilon}+\dfrac{8}{3\,\zeta} \, p(z_{k-1},z_k),&\, \bfX\in \Omega_0
\vspace{0.2cm}\\
\hspace{-0.15cm}\nabla z_k\cdot\bfN=0,& \, \bfX\in \partial\Omega_0
\end{array}\right. \label{BVP-z-theory-reg}
\end{equation}
with
\begin{equation}
p(z_{k-1},z_k)=|z_{k-1}-z_k|-(z_{k-1}-z_k)-|z_k|+z_k,\label{penalty-z}
\end{equation}
$\bfy(\bfX,0)\equiv \bfX$, and $z(\bfX,0)\equiv 1$ for the deformation field $\bfy_k(\bfX)=\bfy(\bfX,t_k)$ and the phase field $z_k(\bfX)=z(\bfX,t_k)$ at any material point $\bfX \in \overline{\Omega}_0=\Omega_0\cup\partial\Omega_0$ and at any given discrete time $t_k\in\{0=t_0,t_1,...,t_m,t_{m+1},$ $...,t_M=T\}$. The penalty parameter $\zeta$ multiplying the regularization function (\ref{penalty-z}) entering equation (\ref{BVP-z-theory-reg})$_1$ should be selected to be small relative to term $2\varepsilon/(\delta^\varepsilon G_c)$. All the results that are presented below correspond to $\zeta^{-1}=10^{4}\delta^\varepsilon G_c/(2\varepsilon)$.

On their own, the governing equations (\ref{BVP-y-theory-reg}) and (\ref{BVP-z-theory-reg}) are second-order PDEs for the deformation field $\bfy_k(\bfX)$ and the phase field $z_k(\bfX)$. Accordingly, their numerical solution is amenable to a FE staggered scheme in which (\ref{BVP-y-theory-reg}) and (\ref{BVP-z-theory-reg}) are discretized with finite elements and solved iteratively one after the other at every time step $t_k$ until convergence is reached. There are two major challenges in the construction of such a scheme. The first one is the selection of an appropriate FE discretization that is capable of dealing with the near incompressibility of the elastomer. The second one is the selection of an appropriate solver for the nonlinear algebraic equations resulting from the discretizations that is capable of dealing with the large changes in the deformation field locally in space and time that can ensue from the nucleation of fracture. As elaborated in Section 4 of \citep{KFLP18}, the first challenge can be efficiently handled by making use of the non-conforming FE discretization of low order due to \cite{CR73}, already mentioned above, while the second challenge can be efficiently handled by making use of an implicit gradient flow solver; see, e.g., \citep{Behrman98,Neuberger10}. All the simulations presented below are generated with such a scheme.

\subsection{The constitutive prescription for the coefficient $\delta^\varepsilon$}\label{Sec: delta}

By construction, the constitutive prescription (\ref{ce-Final})-(\ref{cehat-Final}) for the driving force $c_{\texttt{e}}(\bfX,t)$ leads to a phase-field fracture theory that predicts nucleation of fracture in a specimen that is subjected to uniform stress according to the strength surface (\ref{DP}) of the elastomer, irrespective of the value of the coefficient $\delta^\varepsilon$. However, the value of $\delta^\varepsilon$ \emph{does} affect when the theory predicts nucleation of fracture from a large pre-existing crack, as well as when it predicts propagation of fracture, and hence must be prescribed accordingly.

Following the last step in the construction process for $c_{\texttt{e}}(\bfX,t)$ laid out by \cite{KBFLP20} and \cite{KLP20}, in order to determine the correct value of $\delta^\varepsilon$ for an elastomer with stored-energy function $W(\bfF)$, Drucker-Prager-type strength surface (\ref{DP}) with uniaxial tensile and hydrostatic strengths $s_{\texttt{ts}}$ and  $s_{\texttt{hs}}$, and critical energy release rate $G_c$, for a given regularization length $\varepsilon$, we can consider any boundary-value problem for which the nucleation of fracture from a large pre-existing crack can be determined exactly according to the Griffith energy competition and then have the phase-field theory match that solution thereby determining $\delta^\varepsilon$.

%
\begin{figure}[H]
\begin{center}
\includegraphics[width=0.95\linewidth]{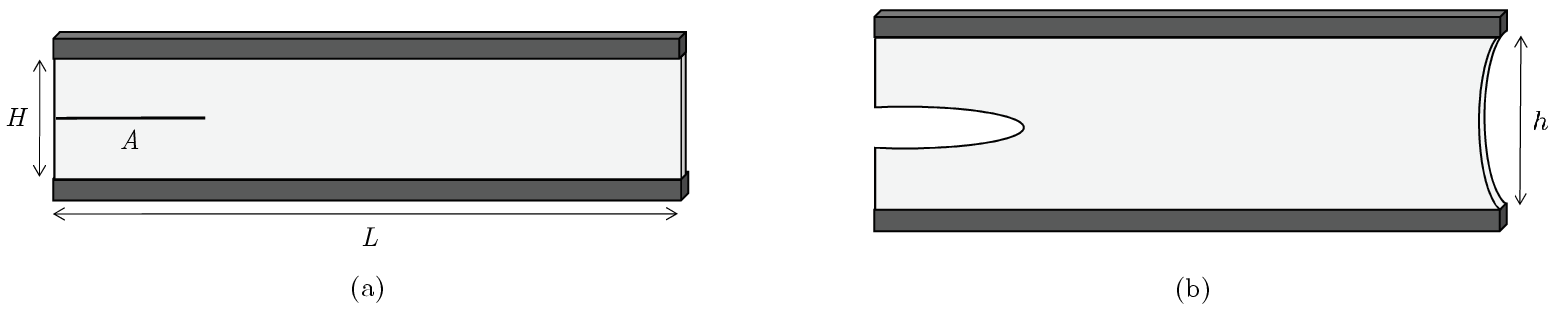}
\end{center}
\vspace{-0.5cm} \caption{Schematic of the ``pure-shear'' fracture test, in (a) the initial configuration and in (b) a deformed configuration at an applied deformation $h$, used to determine the coefficient $\delta^\varepsilon$.}
\label{Fig6}
\end{figure}
%

Among several possible boundary-value problems to choose from, in view of the simplicity of its analysis, a natural choice is the so-called ``pure-shear'' fracture test introduced by \cite{RT53}. As schematically illustrated by Fig. \ref{Fig6}, the specimen in this test consists of a thin sheet of initial height $H$ and much larger length $L\gg H$ that contains a large pre-existing crack, of initial size $A>H$, on one of its sides along its centerline. The specimen is clamped on its top and bottom and subjected to a prescribed deformation $h$ between the grips. For elastomers with stored-energy function (\ref{W-LP}), the critical value $h_{cr}$ of the applied deformation $h$ at which the pre-existing crack will start growing according to the Griffith energy competition can be accurately estimated from the simple equation \citep{RT53}
\begin{equation}
\mathcal{W}(\lambda_{cr}^2+\lambda_{l}^{2}+1,\lambda_{cr}\lambda_{l})=\dfrac{G_c}{H}\quad {\rm with}\quad \lambda_{cr}=\dfrac{h_{cr}}{H},\label{Griffith-PS}
\end{equation}
where the stretch $\lambda_{l}$ is defined implicitly as the root closest to 1 of the nonlinear algebraic equation
\begin{equation*}
0=2\l_{l}\mathcal{W}_{I_1}(\lambda_{cr}^2+\lambda_{l}^{2}+1,\lambda_{cr}\lambda_{l})+\lambda_{cr}\mathcal{W}_{J}(\lambda_{cr}^2+\lambda_{l}^{2}+1,\lambda_{cr}\lambda_{l}).
\end{equation*}
%

%
\begin{figure}[b!]
  \subfigure[]{
   \label{fig:7a}
   \begin{minipage}[]{0.5\textwidth}
   \centering \includegraphics[width=2.75in]{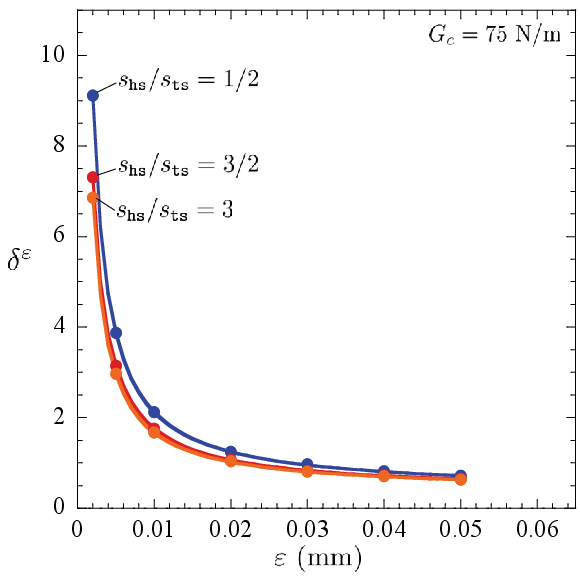}
   \vspace{0.2cm}
   \end{minipage}}
  \subfigure[]{
   \label{fig:7b}
   \begin{minipage}[]{0.5\textwidth}
   \centering \includegraphics[width=2.75in]{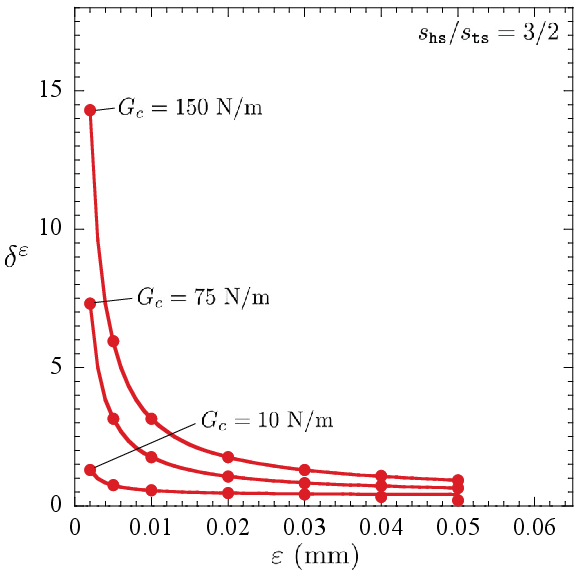}
   \vspace{0.2cm}
   \end{minipage}}
     \subfigure[]{
   \label{fig:7c}
   \begin{minipage}[]{0.5\textwidth}
   \centering \includegraphics[width=2.75in]{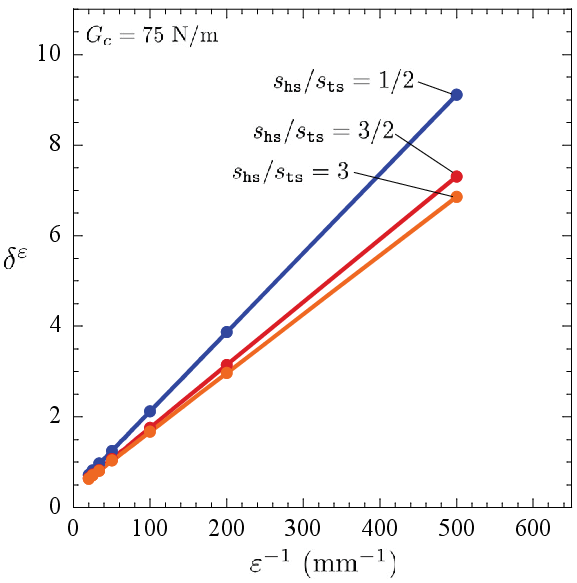}
   \vspace{0.2cm}
   \end{minipage}}
  \subfigure[]{
   \label{fig:7d}
   \begin{minipage}[]{0.5\textwidth}
   \centering \includegraphics[width=2.75in]{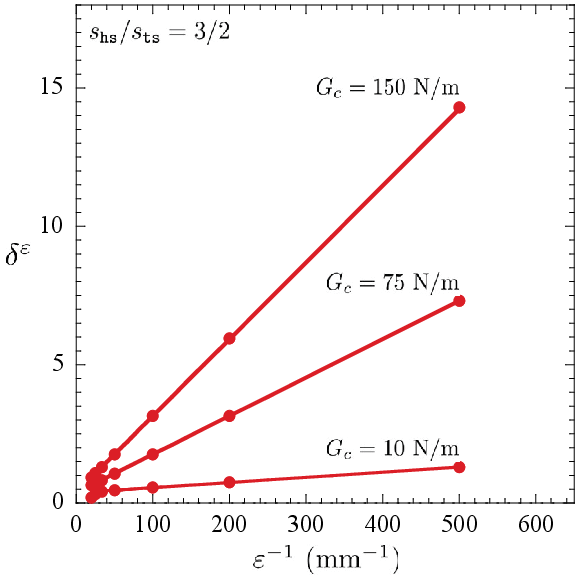}
   \vspace{0.2cm}
   \end{minipage}}
   \caption{Values (solid circles) of the coefficient $\delta^\varepsilon$ determined by matching the theoretical prediction to the Griffith criticality condition (\ref{Griffith-PS}) in a ``pure-shear'' test for an elastomer with stored-energy function (\ref{W-LP}) and the elasticity constants listed in Table \ref{Table1}. The results are shown as functions of the regularization length $\varepsilon$ and of its inverse $\varepsilon^{-1}$ for (a)-(c) $G_c=75$ N/m and the three strength constants $\sts$ and $\shs$ listed in Table \ref{Table1}, so that $s_{\texttt{hs}}/s_{\texttt{ts}}=1/2, 3/2, 3$, and for (b)-(d) $\sts=0.24$ MPa and $\shs=0.36$ MPa, so that $s_{\texttt{hs}}/s_{\texttt{ts}}=3/2$, and the first three critical energy release rates $G_c=10.75,150$ N/m listed in Table \ref{Table1}. For direct comparison, the results generated by the formula (\ref{delta-eps-final-h}) are included (solid lines) in all the plots.}\label{Fig7}
\end{figure}
%

In our simulations, for definiteness, we consider specimens with initial height $H=5$ mm, length $L=50$ mm, and crack size $A=10$ mm. When such specimens are made of the synthetic elastomer of interest here, whose elasticity is characterized by the stored-energy function (\ref{W-LP}) and the material constants listed in Table \ref{Table1}, the values of the critical global stretch $\lambda_{cr}$ at which the crack starts growing according to the Griffith criticality condition (\ref{Griffith-PS}) read
\begin{equation*}
\lambda_{cr}=\left\{\begin{array}{ll}
1.151 & {\rm if}\,\; G_c=10 \; {\rm N/m}\vspace{0.2cm}\\
1.468 & {\rm if}\,\; G_c=75 \; {\rm N/m}\vspace{0.2cm}\\
1.710 &{\rm if}\,\;  G_c=150\; {\rm N/m}\vspace{0.2cm}\\
1.847 &{\rm if}\,\;  G_c=200\; {\rm N/m}\vspace{0.2cm}\\
2.474 &{\rm if}\,\;  G_c=500\; {\rm N/m}\\ \end{array}\right. . 
\end{equation*}

Figure \ref{Fig7} presents the computed values (solid circles) of the coefficient $\delta^\varepsilon$ for which the phase-field theory predicts that the crack in the ``pure-shear'' test starts growing according to (\ref{Griffith-PS}). The results are shown as a function of the regularization length $\varepsilon$ in parts (a) and (b) and as a function of $\varepsilon^{-1}$ in parts (c) and (d). Figures \ref{Fig7}(a) and  \ref{Fig7}(c) present results for an elastomer with critical energy release rate $G_c=75$ N/m and the three pairs of values $(s_{\texttt{ts}},s_{\texttt{hs}})=(0.24,0.12), (0.24,0.36), (0.24,0.72)$ MPa  for the uniaxial tensile and hydrostatic strengths listed in Table \ref{Table1}. On the other hand, Figs. \ref{Fig7}(b) and \ref{Fig7}(d) present results for an elastomer with uniaxial tensile and hydrostatic strengths $\sts=0.24$ MPa and $\shs=0.36$ MPa and the first three values $G_c=10,75,150$ N/m of the critical energy release rate listed in Table \ref{Table1}.

The results in Fig. \ref{Fig7} reveal two key asymptotic aspects of the coefficient $\delta^\varepsilon$, that, to leading order: it increases with the regularization length according to the scaling $\varepsilon^{-1}$ and it depends linearly on the critical energy release rate $G_c$. Precisely, the results reveal that $\delta^\varepsilon$ is of the form

\begin{equation}
\delta^\varepsilon=f_{-1}\left(\mathcal{W}_{\texttt{ts}},\mathcal{W}_{\texttt{hs}},\sts,\shs\right)G_c\varepsilon^{-1}+
f_{0}\left(\mathcal{W}_{\texttt{ts}},\mathcal{W}_{\texttt{hs}},\sts,\shs,G_c\right)+O(\varepsilon),\label{delta-eps-0}
\end{equation}
where the factors $f_{-1}$ and $f_{0}$ are, in principle, functions of all the indicated arguments. The numerical study reported in Appendix A reveals that

\begin{equation}
f_{-1}=\left(\dfrac{\sts+(1+2\sqrt{3})\,\shs}{(8+3\sqrt{3})\,\shs}\right)\dfrac{3}{16\mathcal{W}_{\texttt{ts}}}\qquad {\rm and}\qquad f_{0}=\dfrac{2}{5}\label{factors}
\end{equation}
provide reasonably accurate approximations for these factors. In view of these results, neglecting high-order corrections, we set
\begin{equation}
\delta^\varepsilon=\left(\dfrac{\sts+(1+2\sqrt{3})\,\shs}{(8+3\sqrt{3})\,\shs}\right)\dfrac{3 G_c}{16\mathcal{W}_{\texttt{ts}}\varepsilon}+\dfrac{2}{5}.
\label{delta-eps-final}
\end{equation}

The explicit form (\ref{delta-eps-final}) for the coefficient $\delta^\varepsilon$ is pivotal from both practical and theoretical points of view.
\begin{remark}
\emph{From a practical point of view, the prescription (\ref{delta-eps-final}) makes the driving force (\ref{ce-Final}) fully explicit; this is in contrast to the constitutive prescriptions for $c_{\texttt{e}}(\bfX,t)$ proposed earlier, which required the numerical calibration of a coefficient analogous to $\delta^\varepsilon$; see, e.g., Subsection 4.3.2 in \citep{KBFLP20}. In turn, this makes the governing equations (\ref{BVP-y-theory-reg})-(\ref{BVP-z-theory-reg}) fully explicit in terms of the three material inputs $W(\bfF)$, $\mathcal{F}(\bfS)=0$, and $G_c$ describing the elasticity, strength, and intrinsic fracture energy of the elastomer of interest.}
\end{remark}
\begin{remark}
\emph{From a theoretical point of view, the prescription (\ref{delta-eps-final}) makes the dependence on the regularization length $\varepsilon$ in the evolution equation (\ref{BVP-z-theory-reg}) for the phase field fully explicit. This should allow us to undertake the task of passing to the limit as $\varepsilon\searrow 0$ in the governing equations (\ref{BVP-y-theory-reg})-(\ref{BVP-z-theory-reg}) to determine the corresponding ``sharp'' theory of fracture. We will study such a limit in future work.
}
\end{remark}
\begin{remark}
\emph{When using the FE  method to solve the governing equations (\ref{BVP-y-theory-reg})-(\ref{BVP-z-theory-reg}), meshes of small enough element size $\texttt{h}$ ought to be used so as to appropriately resolve the spatial variations of the phase field $z_k$ over lengths of order $\varepsilon$. Nevertheless, an error is incurred that scales with $\texttt{h}$. Similar to the correction that can be added to $G_c$ within the setting of the original phase-field fracture theory (see, e.g., Section 8.1.1 in \citet{Bourdin08}), here too it is possible to include a similar correction in the formula (\ref{delta-eps-final}) for $\delta^{\varepsilon}$ so that the FE solutions of equations (\ref{BVP-y-theory-reg})-(\ref{BVP-z-theory-reg}) are consistent with the actual value $G_c$ of the critical energy release rate of the elastomer. For first-order finite elements of size $\texttt{h}$, the formula for $\delta^{\varepsilon}$ with the correction reads
\begin{equation}
\delta^\varepsilon=\left(1+\dfrac{3}{8}\dfrac{\texttt{h}}{\varepsilon}\right)^{-2}\left(\dfrac{\sts+(1+2\sqrt{3})\,\shs}{(8+3\sqrt{3})\,\shs}\right)\dfrac{3 G_c}{16\mathcal{W}_{\texttt{ts}}\varepsilon}+\left(1+\dfrac{3}{8}\dfrac{\texttt{h}}{\varepsilon}\right)^{-1}\dfrac{2}{5}.
\label{delta-eps-final-h}
\end{equation}
The results generated by this formula are plotted (solid lines) in Fig. \ref{Fig7} for direct comparison with the data (solid circles) obtained numerically for $\delta^\varepsilon$, which pertains to meshes of element size $\texttt{h}/\varepsilon=1/5$.
}
\end{remark}

\section{Results and discussion}\label{Sec: 2D results}

Having spelled out the governing equations (\ref{BVP-y-theory-reg})-(\ref{BVP-z-theory-reg}), we are now in a position to investigate the deformation and the nucleation and propagation of fracture in poker-chip experiments of synthetic elastomers.

In this section, we carry out a parametric study aimed at identifying the effects that the hydrostatic-to-tensile strength ratio $\shs/\sts$ and the value of the critical energy release rate $G_c$ have on where and when fracture nucleates and propagates in specimens with a range of diameter-to-thickness ratios $D/H$. For computational frugality, since the interest in this section is on the \emph{qualitative} effects of $\shs/\sts$ and $G_c$, we carry out all the simulations under plane-strain conditions, which allows to reduce the problem to a boundary-value problem in 2D. Moreover, we impose symmetry about the centerline of the specimens so that the simulations can be carried out only over one half of the specimens.

Precisely, we consider the governing equations (\ref{BVP-y-theory-reg})-(\ref{BVP-z-theory-reg}) for the deformation field $\bfy(\bfX,t)=y_1(X_1,X_2,t)\bfe_1+y_2(X_2,X_2,t)\bfe_2+X_3\bfe_3$ and phase field $z(\bfX,t)=z(X_1,X_2,t)$ when the specimen occupies the open domain $\Omega_0=\{\bfX:\,|X_1|<D/2,\,|X_2|<H/2\}$ in its initial configuration with respect to the Cartesian frame of reference $\{\bfe_1,\bfe_2\}$. Its top, bottom, and lateral boundaries then read $\partial\Omega^{\mathcal{T}}_0=\{\bfX:\,|X_1|<D/2,\,X_2=H/2\}$, $\partial\Omega^{\mathcal{B}}_0=\{\bfX:\,|X_1|<D/2,\,X_2=-H/2\}$, and $\partial\Omega^{\mathcal{L}}_0=\{\bfX:\,X_1=D/2,\,|X_2|<H/2\}$, respectively. The applied deformations at $\partial\Omega^{\mathcal{T}}_0$ and $\partial\Omega^{\mathcal{B}}_0$ read $\overline{\bfy}^{\mathcal{T}}(\bfX,t)=X_1\bfe_1+(X_2+\dot{h}_0/2\,t)\bfe_2$ and $\overline{\bfy}^{\mathcal{B}}(\bfX,t)=X_1\bfe_1+(X_2-\dot{h}_0/2\,t)\bfe_2$, respectively.

In all results that follow, exactly as in Section \ref{Sec: strength analysis} above, $D=2.5$ cm and $D/H=40,10,4$. The elasticity of the elastomer is characterized by the stored-energy function (\ref{W-LP}) with the material constants listed in Table \ref{Table1}. Its strength is characterized by the Drucker-Prager-type strength surface (\ref{DP}) with uniaxial tensile strength $\sts=0.24$ MPa. To account for the inherently stochastic nature of the strength, we consider a $\pm 10\%$ perturbation of the values  listed in Table \ref{Table1} for the hydrostatic strength $\shs$ that are prescribed randomly over subregions of size $5\varepsilon$ in $\Omega_0$. Finally, we take the critical energy release rate $G_c$ to be given by three of the constant values listed in Table \ref{Table1}. As for the computational parameters, all simulations are carried out with the value $\varepsilon=0.02$ mm for the regularization length making use of unstructured FE meshes of the uniform size $\texttt{h}=\varepsilon/5=0.004$ mm.

\subsection{The effect of the hydrostatic-to-tensile strength ratio $s_{\texttt{\emph{hs}}}/s_{\texttt{\emph{ts}}}$}

\begin{figure}[t!]
\centering
\centering\includegraphics[width=0.92\linewidth]{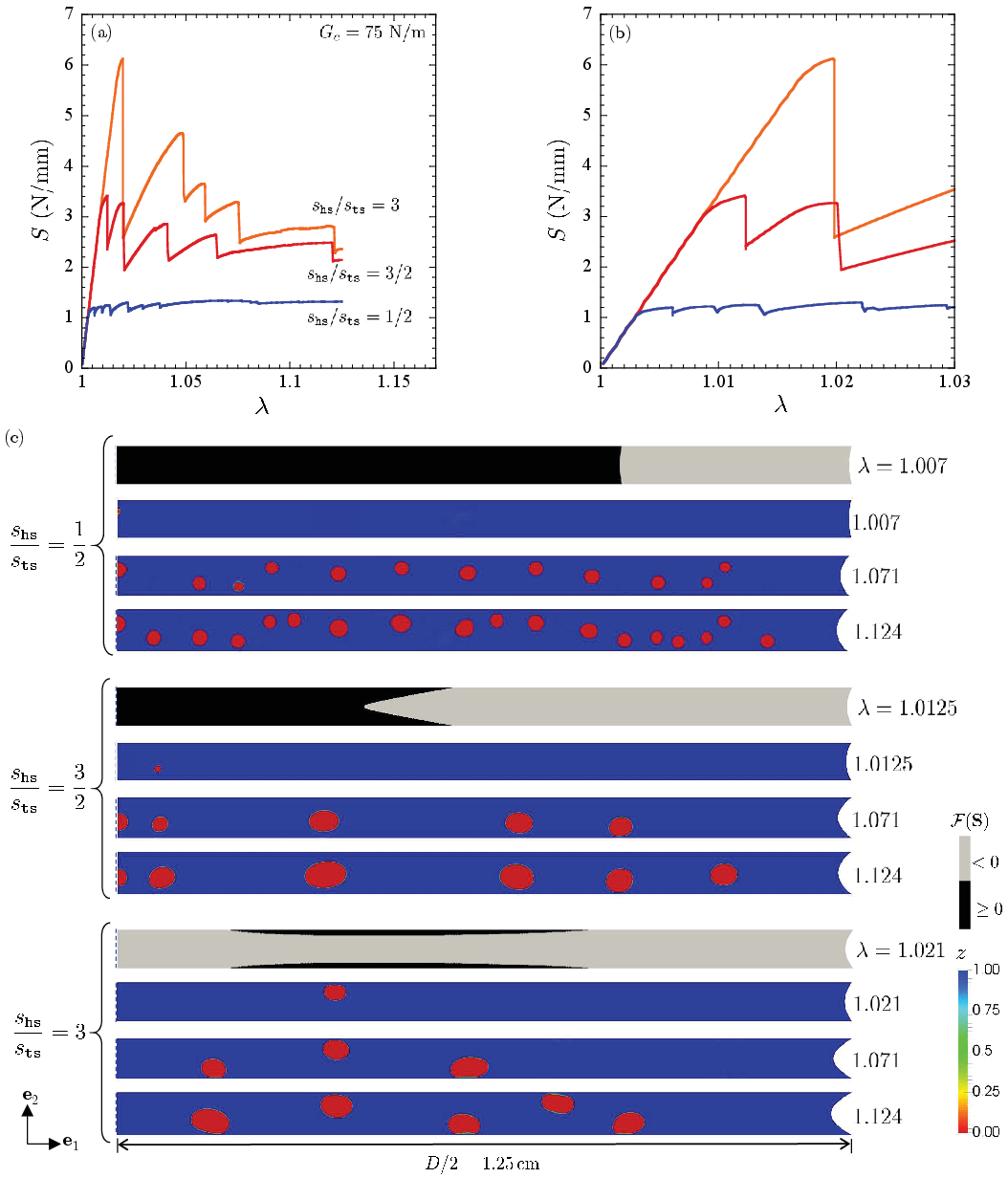}
\caption{\small Simulations of the poker-chip test for specimens with diameter-to-thickness ratio $D/H=40$ made of an elastomer with critical energy release rate $G_c=75$ N/m, uniaxial tensile strength $s_{\texttt{ts}}=0.24$ MPa, and the three different values of hydrostatic strength $s_{\texttt{hs}}$ listed in Table \ref{Table1}, so that $s_{\texttt{hs}}/s_{\texttt{ts}}=1/2, 3/2, 3$. (a) The normalized force $S=P/D$ as a function of the normalized global deformation $\lambda=h/H$. (b) Close-up of the $S$-$\lambda$ response in the small-deformation region. (c) Contour plots, over the deformed configuration, of the phase field $z$ and of the regions that violate the strength surface of the elastomer at select values of the applied deformation $\lambda$. The contour plots are shown only over half of the specimens for better visualization.}\label{Fig8}
\end{figure}
\begin{figure}[t!]
\centering
\centering\includegraphics[width=0.9\linewidth]{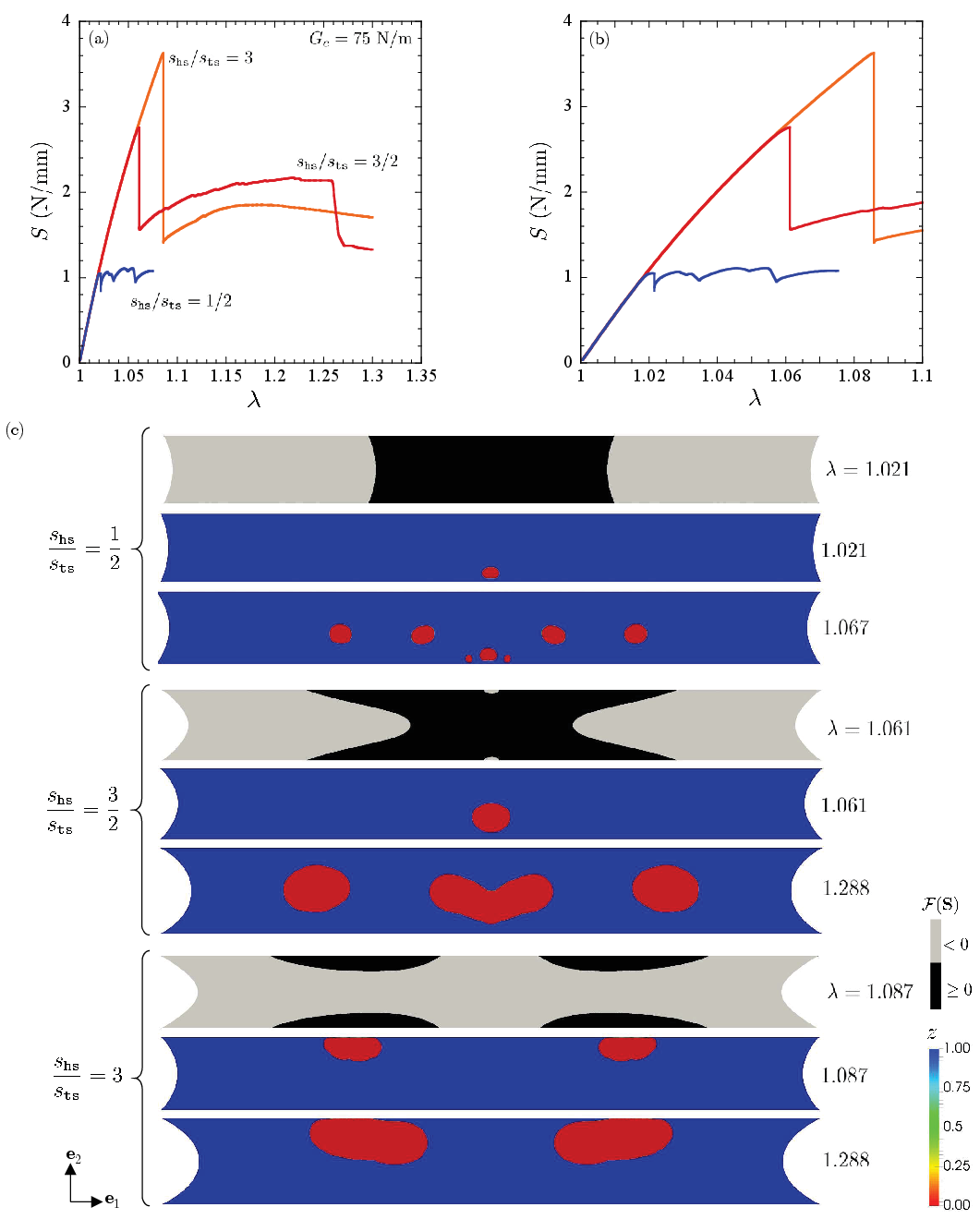}
\caption{\small Simulations of the poker-chip test for specimens with diameter-to-thickness ratio $D/H=10$ made of an elastomer with critical energy release rate $G_c=75$ N/m, uniaxial tensile strength $s_{\texttt{ts}}=0.24$ MPa, and the three different values of hydrostatic strength $s_{\texttt{hs}}$ listed in Table \ref{Table1}, so that $s_{\texttt{hs}}/s_{\texttt{ts}}=1/2, 3/2, 3$. (a) The normalized force $S=P/D$ as a function of the normalized global deformation $\lambda=h/H$. (b) Close-up of the $S$-$\lambda$ response in the small-deformation region. (c) Contour plots, over the deformed configuration, of the phase field $z$ and of the regions that violate the strength surface of the elastomer at select values of the applied deformation $\lambda$.}\label{Fig9}
\end{figure}
\begin{figure}[t!]
\centering
\centering\includegraphics[width=0.9\linewidth]{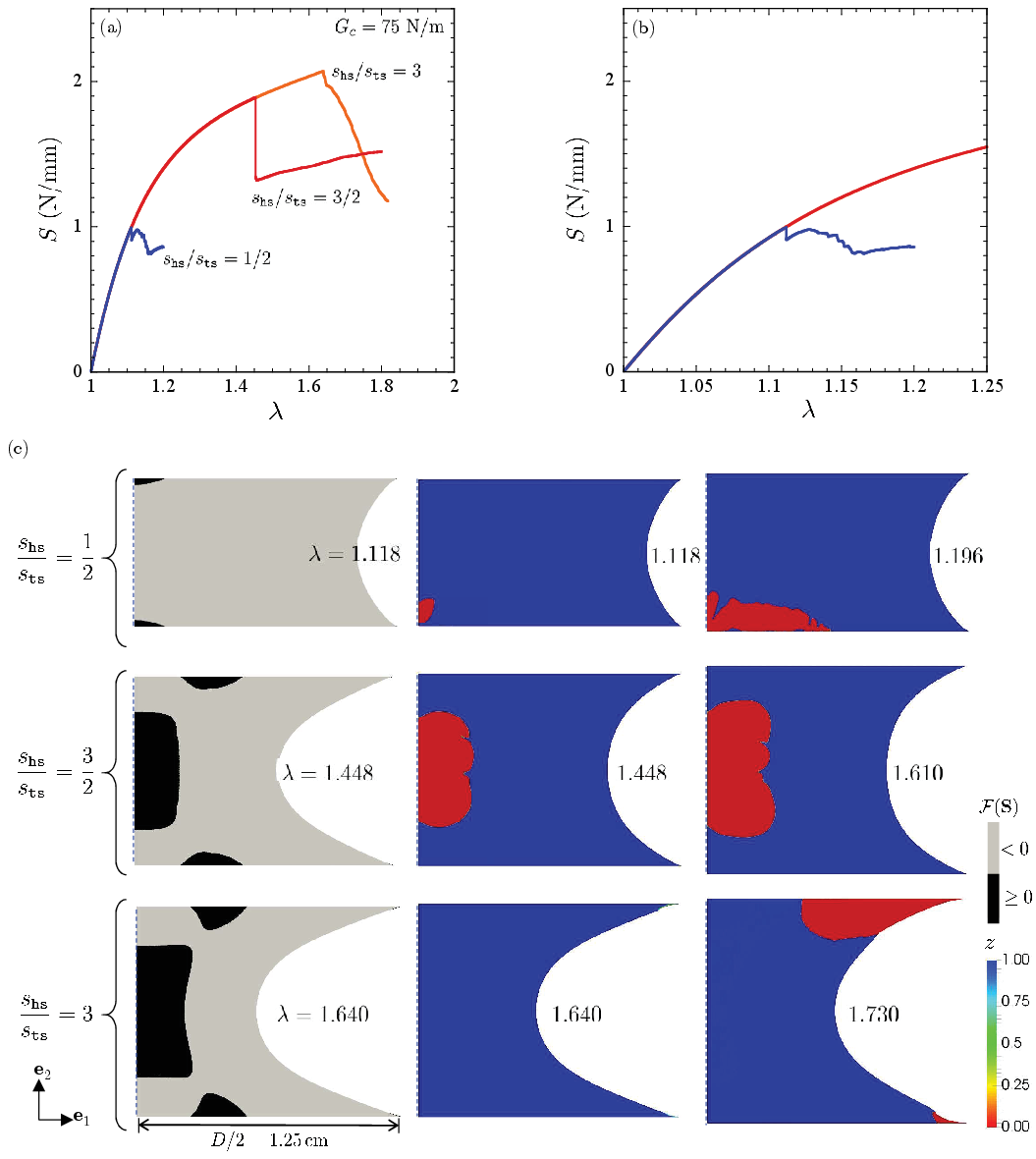}
\caption{\small Simulations of the poker-chip test for specimens with diameter-to-thickness ratio $D/H=4$ made of an elastomer with critical energy release rate $G_c=75$ N/m, uniaxial tensile strength $s_{\texttt{ts}}=0.24$ MPa, and the three different values of hydrostatic strength $s_{\texttt{hs}}$ listed in Table \ref{Table1}, so that $s_{\texttt{hs}}/s_{\texttt{ts}}=1/2, 3/2, 3$. (a) The normalized force $S=P/D$ as a function of the normalized global deformation $\lambda=h/H$. (b) Close-up of the $S$-$\lambda$ response in the small-deformation region. (c) Contour plots, over the deformed configuration, of the phase field $z$ and of the regions that violate the strength surface of the elastomer at select values of the applied deformation $\lambda$. The contour plots are shown only over half of the specimens for better visualization.}\label{Fig10}
\end{figure}
\begin{figure}[t!]
\centering
\centering\includegraphics[width=0.92\linewidth]{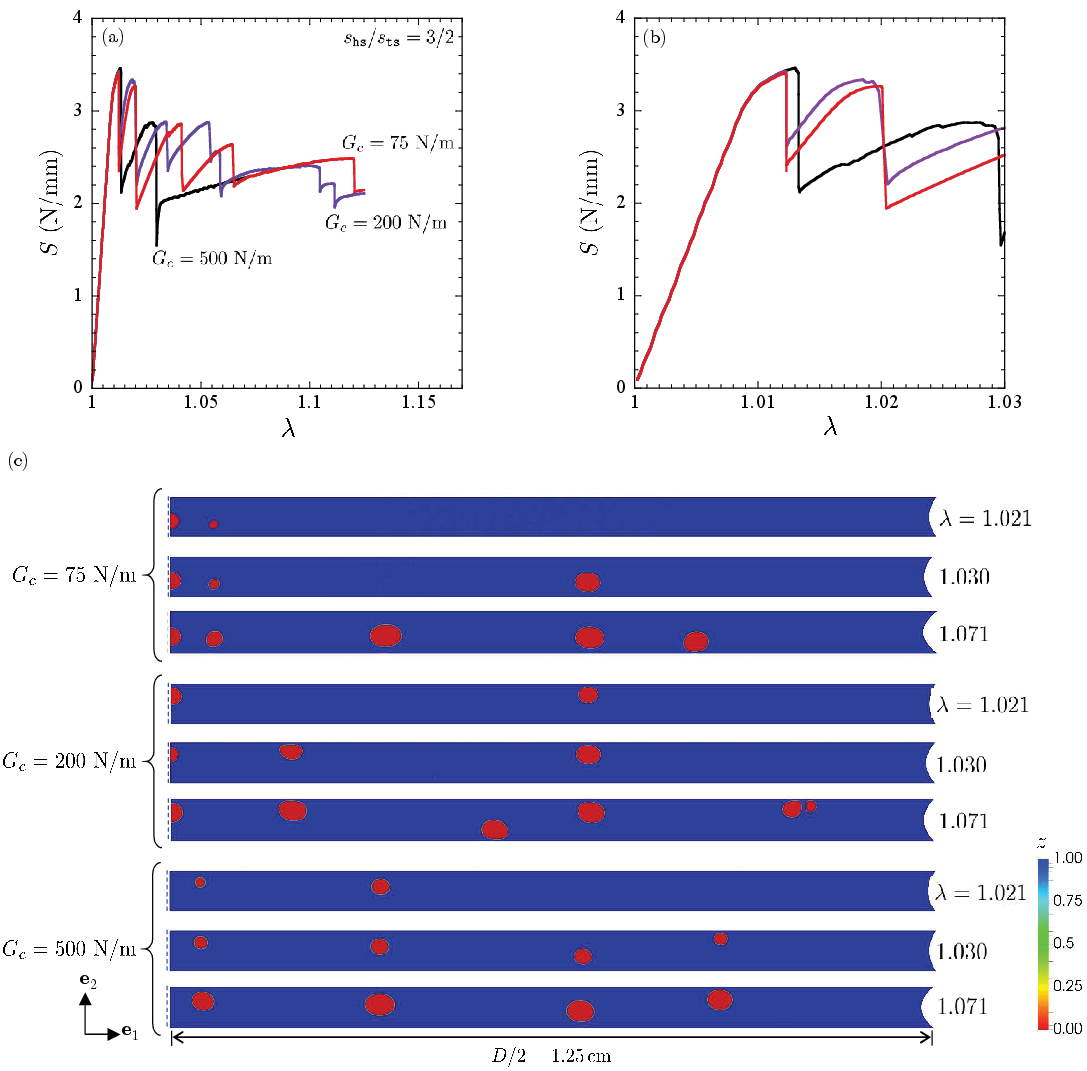}
\caption{\small Simulations of the poker-chip test for specimens with diameter-to-thickness ratio $D/H=40$ made of an elastomer with uniaxial tensile strength $\sts=0.24$ MPa, hydrostatic strength $\shs=0.36$ MPa, and three of the different values of critical energy release rate $G_c$ listed in Table \ref{Table1}. (a) The normalized force $S=P/D$ as a function of the normalized global deformation $\lambda=h/H$. (b) Close-up of the $S$-$\lambda$ response in the small-deformation region. (c) Contour plots, over the deformed configuration, of the phase field $z$ at select values of the applied deformation $\lambda$. The contour plots are shown only over half of the specimens for better visualization.}\label{Fig11}
\end{figure}
\begin{figure}[t!]
\centering
\centering\includegraphics[width=0.92\linewidth]{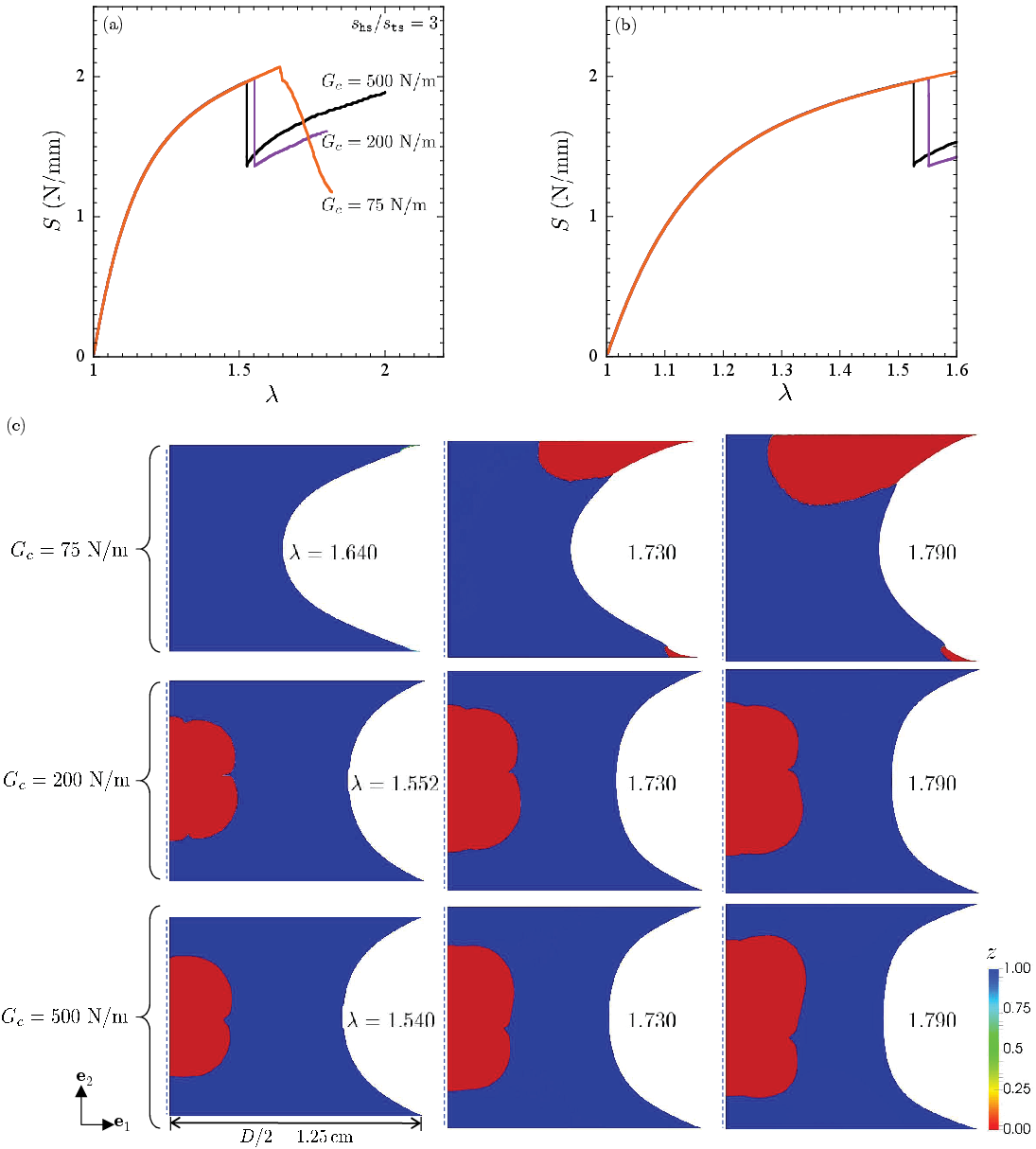}
\caption{\small Simulations of the poker-chip test for specimens with diameter-to-thickness ratio $D/H=4$ made of an elastomer with uniaxial tensile strength $\sts=0.24$ MPa, hydrostatic strength $\shs=0.72$ MPa, and three of the different values of critical energy release rate $G_c$ listed in Table \ref{Table1}. (a) The normalized force $S=P/D$ as a function of the normalized global deformation $\lambda=h/H$. (b) Close-up of the $S$-$\lambda$ response in the small-deformation region. (c) Contour plots, over the deformed configuration, of the phase field $z$ at select values of the applied deformation $\lambda$. The contour plots are shown only over half of the specimens for better visualization.}\label{Fig12}
\end{figure}

Figure \ref{Fig8} presents results from the simulation of a poker-chip test for very thin specimens with diameter-to-thickness ratio $D/H=40$. The results pertain to an elastomer with critical energy release rate $G_c=75$ N/m and the three values $s_{\texttt{hs}}=0.12, 0.36, 0.72$ of hydrostatic strength so that $s_{\texttt{hs}}/\sts=1/2, 3/2, 3$. Specifically, Figs. \ref{Fig8}(a)-(b) show the normalized global force $S=P/D$ as a function of the normalized global deformation $\lambda=h/H$; whereas part (a) shows the results in the entire range of applied deformations $\lambda\in[1,1.12]$, part (b) zooms in the small-deformation region $\lambda\in[1,1.03]$. Figure \ref{Fig8}(c) shows accompanying contour plots, over the deformed configuration, of the phase field $z(\bfX,t)$ at three select applied deformations $\lambda$, corresponding to the first nucleation of fracture and to two subsequent states showing additional fracture nucleation events and fracture propagation. To aid the discussion, Fig. \ref{Fig8}(c) also includes the contour plots of the regions that violate the strength surface of the elastomer at the first fracture nucleation event.

There are several observations from Figs. \ref{Fig8}(a)-(b) that are noteworthy. First, the force-deformation response of the specimens exhibits two distinct successive regions: a stiff, approximately linear, response from $\lambda=1$ up to some critical value $\lambda>1$ beyond which the response transitions to a rugged plateau. Since all three specimens are made of an elastomer with the same elasticity, the slope of the initial stiff response is the same for all three. The critical values of the applied deformation $\lambda$ at which the response transitions to a rugged plateau, however, are larger for the elastomers with larger hydrostatic-to-tensile strength ratio $\shs/\sts$. The ruggedness of the plateau is also different. For increasing values of the strength ratio $\shs/\sts$, the frequency in $\lambda$ of local maxima and minima decreases, but the difference in force values $S$ between the local maxima and minima increases.  

The results in Figs. \ref{Fig8}(a)-(b) can be easily understood by turning to the contour plots in Fig. \ref{Fig8}(c). The latter reveal that the initial stiff force-deformation response of the specimens is due solely to the elastic deformation of the elastomer and that the transition to the rugged plateau is nothing more than the manifestation of the first nucleation of cracks. For the elastomer with the smallest strength ratio $\shs/\sts=1/2$, a single crack first nucleates along the centerline of the specimen. On the other hand, for the elastomers with the larger strength ratios $\shs/\sts=3/2$ and $3$, cracks first nucleate \emph{radially away from the centerline}. For all three elastomers, as anticipated in Remark \ref{Remark-bound} above, the contour plots illustrating the regions where the strength surface has been violated corroborates that the violation of the strength surface is a necessary but \emph{not} sufficient condition for nucleation of fracture.

Upon further loading, irrespective of the strength ratio $\shs/\sts$, the nucleated cracks exhibit limited propagation, instead, they mostly deform and more cracks are nucleated at adjacent locations. The nucleation of new adjacent cracks together with the limited propagation of all previously nucleated cracks continues until the greater part of the specimens is substantially populated with cracks. The rugged plateau region in the force-deformation response is precisely the result of this cascading process. Another immediate observation is that the number of cracks that are nucleated increases with decreasing $\shs/\sts$. This is the reason why the frequency in $\lambda$ of the ruggedness of the plateau is different for different values of the strength ratio $\shs/\sts$.

Figure \ref{Fig9} presents results analogous to those presented in Fig. \ref{Fig8} for thin specimens with diameter-to-thickness ratio $D/H=10$. The force-deformation response in Figs. \ref{Fig9}(a)-(b) shows the same two distinct regions shown by the very thin poker-chip specimens in Figs. \ref{Fig8}(a)-(b). There are, however, several differences. The initial response is less stiff and slightly nonlinear. The transition to the rugged plateau region occurs at larger critical values of the applied  deformation $\lambda$. Finally, the frequency in $\lambda$ of the ruggedness of the plateau is less significant.

The force-deformation response in Figs. \ref{Fig9}(a)-(b) can too be understood at once from the corresponding contour plots of the phase field in Fig. \ref{Fig9}(c). The latter reveal that the initial part of the force-deformation response of the specimens is due solely to the elastic deformation of the elastomer and that the transition to the rugged plateau is, again, the manifestation of the first nucleation of cracks. For the elastomers with the two smaller strength ratios $\shs/\sts=1/2$ and $3/2$, a single crack first nucleates along the centerline of the specimens near one of the fixtures. On the other hand, for the elastomer with the largest strength ratio $\shs/\sts=3$, cracks first nucleate \emph{radially away from the centerline}, roughly, at a distance $D/4$ away, also near one of the fixtures.

Similar to the behavior observed for the very thin specimens, irrespective of the strength ratio $\shs/\sts$, the nucleated cracks exhibit limited propagation upon further loading, more so the smaller the value of $\shs/\sts$. They mostly deform and, for the elastomers with the two smaller strength ratios $\shs/\sts=1/2$ and $3/2$ but not for the one with the largest strength ratio $\shs/\sts=3$, more cracks are nucleated at adjacent locations. The number of cracks that are nucleated in these thin specimens is substantially smaller than the number of cracks that are nucleated in the very thin specimens. Similar to the behavior observed for the very thin specimens, however, the number of cracks that are nucleated increases with decreasing $\shs/\sts$.

Figure \ref{Fig10} presents results analogous to those presented in Figs. \ref{Fig8} and \ref{Fig9}, now for thick specimens with diameter-to-thickness ratio $D/H=4$. The force-deformation response in Figs. \ref{Fig10}(a)-(b) appears to show the same two distinct regions shown by the very thin and the thin poker-chip specimens in Figs. \ref{Fig8}(a)-(b) and Figs. \ref{Fig9}(a)-(b), albeit with significant quantitative differences. Specifically, the initial response is less stiff and distinctly nonlinear and the transition to the rugged plateau region, or the sharp softening for the elastomer with the largest strength ratio $\shs/\sts=3$, occurs at larger critical values of the applied  deformation $\lambda$.

The force-deformation response of the thick specimens in Figs. \ref{Fig10}(a)-(b) can be understood yet again from the corresponding contour plots of the phase field in Fig. \ref{Fig10}(c). These contour plots show that the initial stiff response of the specimens is due solely to the elastic deformation of the elastomer and that the transition to the rugged plateau corresponds to the first nucleation of cracks. For the elastomer with the largest strength ratio $\shs/\sts=3$, the first nucleation of a crack occurs at the free boundary of the specimen, next to one of the fixtures. On the other hand, for the elastomer with the strength ratio $\shs/\sts=1/2$, a single crack first nucleates along the centerline of the specimens near one of the fixtures, while for the elastomer with $\shs/\sts=3/2$, a single crack first nucleates around the center of the specimen.

Upon further loading, irrespective of the strength ratio $\shs/\sts$, the nucleated cracks deform and propagate. For the elastomer with the largest strength ratio $\shs/\sts=3$, the propagation of the crack leads to a sharp softening in the force-deformation response of the specimen, during which a second crack nucleates from the free boundary of the specimen, next to the opposite fixture.

\subsection{The effect of the critical energy release rate $G_c$}

We now turn to probing the effect of the value of the critical energy release rate $G_c$. Figure \ref{Fig11} presents results from the simulation of a poker-chip test for very thin specimens with diameter-to-thickness ratio $D/H=40$. The results pertain to an elastomer with uniaxial tensile strength $\sts=0.24$ MPa, hydrostatic strength $\shs=0.36$ MPa,  and the three values of critical energy release rate $G_c=75, 200, 500$ N/m. Analogous to the preceding results, Figs. \ref{Fig11}(a)-(b) show the normalized global force $S=P/D$ as a function of the normalized global deformation $\lambda=h/H$, while Fig. \ref{Fig11}(c) shows accompanying contour plots, over the deformed configuration, of the phase field $z(\bfX,t)$ at three select applied deformations $\lambda$, corresponding to the first nucleation of fracture and to two subsequent states showing additional fracture nucleation events and fracture propagation.

As the main observation, it is plain from  Fig. \ref{Fig11} that the critical energy release rate $G_c$ --- so long as is constant and its value is within the realistic range for synthetic elastomers --- has a minor effect on how fracture nucleates and propagates in very thin poker-chip specimens. In other words, the value of $G_c$ has a minor effect on the behavior of poker-chip specimens where there are multiple fracture nucleation events.

On the other hand, the value of $G_c$ has a major effect on how fracture nucleates and propagates in poker-chip experiments where there is just one or a few fracture nucleation events. This is clearly illustrated by Fig. \ref{Fig12}, where results analogous to those in Fig. \ref{Fig11} are presented for thick specimens with diameter-to-thickness ratio $D/H=4$ made of an elastomer with uniaxial tensile strength $\sts=0.24$ MPa, hydrostatic strength $\shs=0.72$ MPa, and the three critical energy release rates $G_c=75, 200, 500$ N/m.

The results in Fig. \ref{Fig12} show that fracture nucleation in elastomers with a large hydrostatic strength $\shs$ relative to the uniaxial tensile strength $\sts$ and a sufficiently small critical energy release rate $G_c$  ($75$  N/m in this case) can occur at the free boundary of thick specimens, near one of the fixtures. Upon further loading, that crack propagates. However, for elastomers with a sufficiently large $G_c$ ($200$ and $500$ N/m in this case), no crack is nucleated near the free boundary, instead, a crack nucleates along the centerline of the specimen and, upon further loading, it propagates. In other words, a sufficiently large value of $G_c$ can preclude the nucleation of cracks from the free boundary in thick poker-chip specimens.

\subsection{Summary of the results and how they explain experimental observations on synthetic elastomers}

\subsubsection{Nucleation}\label{Sec: Nucleation}

Summing up, the simulations presented above have established that the nucleation of internal cracks in a poker-chip experiment occurs in regions where the strength surface of the elastomer has been exceeded, soon after it has been exceeded. This is because the stress field in a poker-chip specimen is not excessively non-uniform --- save for the singular stress at the curves where the free boundary meets the fixtures --- and, hence, nucleation of fracture in the interior of the specimen is dominated primarily by the strength of the elastomer.

On the other hand, the simulations have also shown that the nucleation of cracks from the free boundary of the specimen occurs from the geometric singular curves where the free boundary meets the fixtures as dictated by an ``interpolation'' between the strength of the elastomer and its critical energy release rate. This ``interpolation'' is implicit to the governing equations (\ref{BVP-y-theory-reg}) and (\ref{BVP-z-theory-reg}). As alluded to in Remark \ref{Remark_length} above and illustrated in Appendix B below, it is a direct consequence of the family of material length scales that is intrinsic to the theory and that comes about because the evolution equation (\ref{BVP-z-theory-reg}) for the phase field depends on material inputs of different units. Precisely, the elastic stored-energy function $W(\bfF)$ and the strength surface $\mathcal{F}(\bfS)=0$ have units of $force/length^2$, while the critical energy release rate $G_c$ has units of $force/length$. Their combination in equation (\ref{BVP-z-theory-reg}) leads to the family of material length scales in the theory.

Accordingly, the first nucleation of fracture can occur either around the centerline of the specimen or radially away from the centerline. The precise location where fracture first nucleates depends critically on the hydrostatic-to-tensile strength ratio $\shs/\sts$ of the elastomer, its critical energy release rate $G_c$, as well as, of course, on the diameter-to-thickness ratio $D/H$ of the specimen.

For elastomers for which the hydrostatic strength $\shs$ is small relative to the uniaxial tensile strength $\sts$ --- which, again, it is case of natural rubber ---  the first nucleation of fracture always occurs along the centerline of the specimen, near one of the fixtures, this irrespective of the values of $G_c$ and $D/H$. This result was already established by the analysis of \cite{KLP21} of the classical poker-chip experiments of \cite{GL59} on natural rubber.

For elastomers for which the hydrostatic strength $\shs$ is comparable to or larger than the uniaxial tensile strength $\sts$, the first nucleation of fracture can occur either along the centerline of the specimen or radially away from it depending on the critical energy release rate $G_c$ of the elastomer and on the diameter-to-thickness ratio $D/H$ of the specimen. These results serve to explain recent observations from poker-chip experiments on synthetic elastomers, in particular, the experiments of \cite{GuoRavi23} on PDMS 30:1, where the first nucleation of cracks was observed to occur substantially away from the centerline of the specimens\footnote{The fact that internal cracks can first nucleate radially away from the centerline of poker-chip specimens is one of the fundamental evidence that rule out cavitation being a purely elastic phenomenon \citep{BCLP24}.} as well as from the free boundary, in stark contrast to the observations by \cite{GL59} on natural rubber.

\begin{figure}[t!]
\centering
\centering\includegraphics[width=0.41\linewidth]{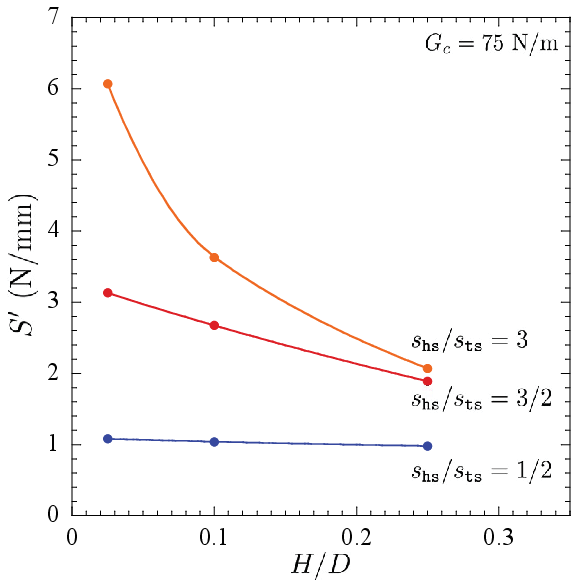}
\caption{\small The critical value $S^\prime$ of the normalized global force $S=P/D$ at which nucleation of cracks first occurs in poker-chip tests as a function of the thickness $H$ of the specimen, normalized by its diameter $D$. The results correspond to an elastomer with critical energy release rate $G_c=75$ N/m, uniaxial tensile strength $s_{\texttt{ts}}=0.24$ MPa, and the three different values of hydrostatic strength $s_{\texttt{hs}}$ listed in Table \ref{Table1}, so that $s_{\texttt{hs}}/s_{\texttt{ts}}=1/2, 3/2, 3$.}\label{Fig13}
\end{figure}

Larger hydrostatic-to-tensile strength ratios $\shs/\sts$ also lead to larger critical values $S^\prime$ of the normalized global force $S$ at which the first nucleation of fracture occurs. This is better illustrated by Fig. \ref{Fig13}, where $S^\prime$ is collected from the results in Figs. \ref{Fig8} through \ref{Fig10} above and plotted as a function of the normalized thickness $H/D$ of the specimen. Figure \ref{Fig13} serves to illustrate as well that $S^\prime$ decreases with increasing specimen thickness $H/D$, more so the larger the hydrostatic-to-tensile strength ratio $\shs/\sts$, in agreement with experimental observations; cf., Fig. 6 in \citep{GL59} and Fig. 24 in \citep{GuoRavi23}.

The simulations have served to reveal too that larger hydrostatic-to-tensile strength ratios $\shs/\sts$ lead to a smaller total number of cracks that are nucleated. This is because larger $\shs$ relative to $G_c$ favors the propagation of previously nucleated cracks in detriment of the nucleation of new cracks. Accordingly, an increase in the strength ratio $\shs/\sts$ of the elastomer has a similar effect as a decrease in the diameter-to-thickness ratio $D/H$ of the specimen.  

\subsubsection{Propagation}

Interestingly, the simulations have shown that the precise value of the critical energy release rate $G_c$ of synthetic elastomers (which is bounded from above by a few hundreds of N/m) has a minor effect on how cracks propagate in thin poker-chip specimens, provided that $\shs$ and $\sts$  are sufficiently small.

For sufficiently thick poker-chip specimens, on the other hand, when just one or a few cracks are nucleated, the value of $G_c$ can have a major effect leading to the arrest of some cracks and the propagation of others, depending on their location in the specimen, if $G_c$ is sufficiently large.

\section{Comparison with the experiments of Guo and Ravi-Chandar (2023) on PDMS 30:1}\label{Sec: Experiments}

The parametric study presented in the preceding section has shown that the hydrostatic-to-tensile strength ratio $\shs/\sts$ and the value of the critical energy release rate $G_c$ have a profound impact on where and when cracks nucleate and propagate in poker-chip experiments of synthetic elastomers. By the same token, they have shown that the difference in these two material properties with respect to those of natural rubber is precisely the reason why poker-chip experiments of synthetic elastomers exhibit different behaviors from those observed by \cite{GL59}  on natural rubber. In this section, to illustrate this key difference between synthetic elastomers and natural rubber in a quantitative manner that complements the qualitative parametric study of Section \ref{Sec: 2D results} above, we present simulations of the recent poker-chip experiments carried out by \cite{GuoRavi23} on PDMS 30:1 and compare them directly to their measurements and observations. In particular, we present simulations for the tests that \cite{GuoRavi23} referred to as Tests D, C, and A.

\subsection{The geometry of the specimens}\label{Sec: Geometry}

\cite{GuoRavi23} reported that their Tests D, C, and A pertain to specimens of diameter $D=2.5$ cm and diameter-to-thickness ratios
\begin{equation*}
\dfrac{D}{H}=41.6, 11.3, 4.7,
\end{equation*}
respectively. These values correspond to initial thicknesses
\begin{equation*}
H=0.60, 2.21, 5.32\, {\rm mm}.
\end{equation*}
The spatial resolution in these measurements of $H$ was not reported. Based on our simulations, we estimate it to be around $\pm0.1$ mm. This resolution has little impact on the thin ($H=2.21\pm 0.1$ mm) and thick ($H=5.32\pm 0.1$ mm) specimens, but it does have a significant impact on the very thin ($H=0.60\pm 0.1$ mm) one. In the simulations that follow, we assume that the initial thicknesses of the specimens used in Tests D, C, and A by \cite{GuoRavi23} are in fact
\begin{equation*}
H=0.50, 2.21, 5.32\, {\rm mm},
\end{equation*}
respectively. These correspond to specimens with diameter-to-thickness ratios
\begin{equation*}
\dfrac{D}{H}=50, 11.3, 4.7.
\end{equation*}

\subsection{Calibration of the three material inputs entering the theory}\label{Sec: calibration}

\paragraph{The elasticity} As noted in Subsection \ref{Sec: Elasticity} above, \cite{Poulain17} showed that the elastic behavior of PDMS Sylgard 184 with a 30:1 weight ratio of base elastomer to cross-linking agent is well described by the stored-energy function (\ref{W-LP}) with the material constants $\mu_r$, $\alpha_r$, and $\kappa$ listed in Table \ref{Table1}. In all the simulations presented below, we make use of that constitutive relation for the elasticity of PDMS 30:1.

\paragraph{The strength} As also already alluded to in Subsection \ref{Sec: Strength} above, \cite{KLP20} estimated from experiments of \cite{Poulain17} that the uniaxial tensile strength and the hydrostatic strength of PDMS 30:1 are around $\sts=0.2$ MPa and $\shs=0.5$ MPa. In all the simulations presented below, we take the strength of PDMS 30:1 to be described by the Drucker-Prager-type strength surface (\ref{DP}) with the following deterministic uniaxial tensile strength and stochastic hydrostatic strength:
\begin{equation}
\sts=0.24\, {\rm MPa}\quad {\rm and}\quad \shs\in[0.64,0.80]\,{\rm MPa}.\label{sts-shs-PDMS}
\end{equation}

Note that the range of values (\ref{sts-shs-PDMS})$_2$ for the hydrostatic strength corresponds to an average value $\shs=0.72$ MPa with a $\pm 10\%$ variation. In the simulations, this variation in $\shs$ is prescribed randomly over subregions of size $5\varepsilon$ in the initial domain $\Omega_0$ occupied by the specimens. The value (\ref{sts-shs-PDMS})$_1$ for the uniaxial tensile strength is kept to be determinist for simplicity. Making $\sts$ stochastic does not have a significant effect on the results, since the stochasticity in $\shs$ already generates stochasticity in the entire strength surface (\ref{DP}).

Note also that the strength constants (\ref{sts-shs-PDMS}) correspond to the range of hydrostatic-to-tensile strength ratios
\begin{equation*}
\dfrac{\shs}{\sts}\in[2.67,3.33],
\end{equation*}
which, again, is drastically different from that $\shs/\sts=0.24$ of natural rubber \citep{KLP21}.

\paragraph{The critical energy release rate} \cite{GuoRavi23} did not report measurements for the critical energy release rate $G_c$ of PDMS 30:1. As noted in Subsection \ref{Sec: Gc} above, nonetheless, experiments on similar silicone elastomers \citep{GT82} suggest that the critical energy release rate of PDMS 30:1 should be in the ballpark of $G_c=75$ N/m. In all the simulations presented below, we make use of the value $G_c=75$ N/m.

\subsection{Computational aspects}

To reduce the significant computational cost associated with the simulations, we assume that the solution is symmetric per an \texttt{oct}ant of the specimen, which allows to carry out the computations over the initial subdomain
\begin{equation}\label{Oct}
\Omega^{\texttt{Oct}}_0=\left\{\bfX:\sqrt{X_1^2+X_2^2}< \dfrac{D}{2},\, X_3< \dfrac{H}{2},\,X_1,X_2,X_3>0\right\}
\end{equation}
by imposing the symmetry conditions $y_1(0,X_2,X_3)=0$, $y_2(X_1,0,X_3)=0$, $y_3(X_1,X_2,0)=0$ on the deformation field $\bfy(\bfX,t)$  at the planes of symmetry; see Fig. \ref{Fig1}.

In the simulation for the very thin poker-chip experiment with diameter-to-thickness ratio $D/H=50$, we make use of the regularization length $\varepsilon=0.01$ mm. In the simulations for the thin and thick specimens with $D/H=11.3$ and $4.7$, on the other hand, we make use of the larger value $\varepsilon=0.02$ mm. All simulations are carried out with unstructured FE meshes of the uniform size $\texttt{h}=\varepsilon/5$.

\subsection{Theory vs. experiment}\label{Sec:force-def-very-thin}

\begin{figure}[t!]
\centering
\centering\includegraphics[width=0.89\linewidth]{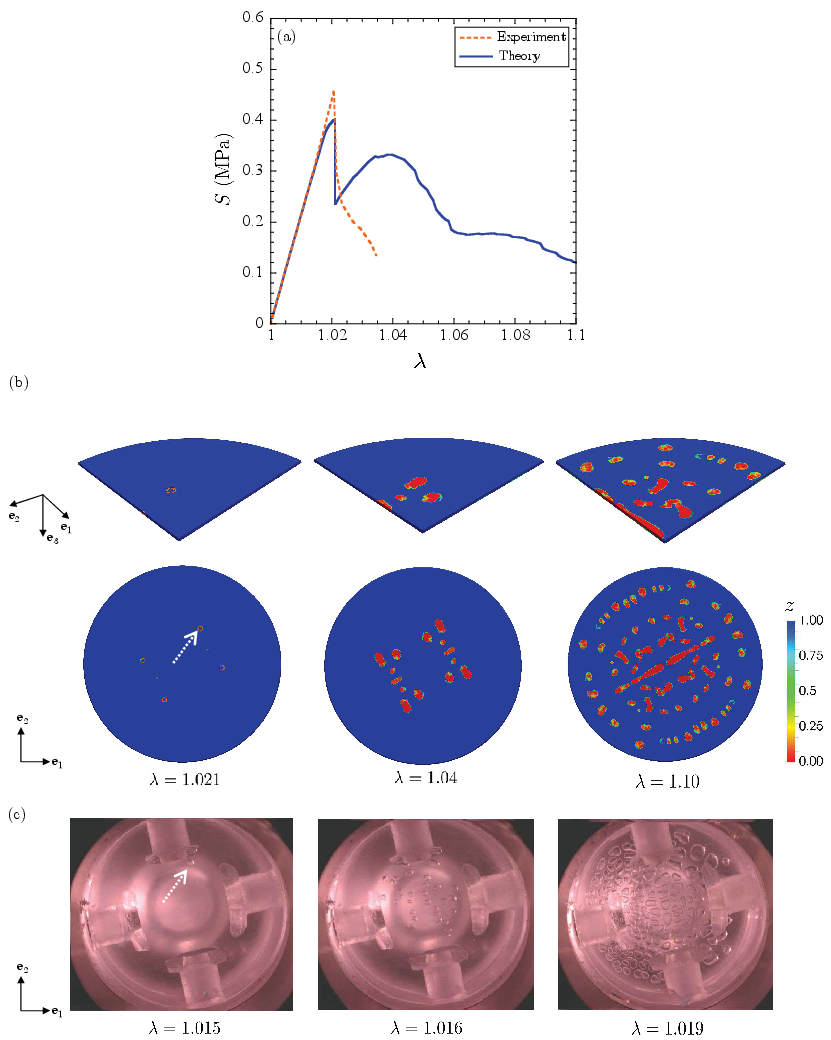}
\caption{\small Comparison between theory and experiment for the very thin poker-chip specimen with diameter-to-thickness ratio $D/H=50$. (a) The normalized force $S=4P/(\pi D^2)$ as a function of the normalized applied deformation $\lambda=h/H$. (b) Contour plots, over the deformed configuration, of the phase field $z$ at three select applied deformations $\lambda$, as predicted by the theory. (c) Selected sequence of images from the experiment at three different values of $\lambda$. The white arrows in parts (b) and (c) identify the first nucleated crack. \vspace{0.2cm}}\label{Fig14}
\end{figure}

Figure \ref{Fig14} confronts the results predicted by the theory with the experimental data for the very thin poker-chip experiment with diameter-to-thickness ratio $D/H=50$; see Fig. 8 in \citep{GuoRavi23}. Specifically, Fig. \ref{Fig14}(a) compares the force-deformation responses, while Figs. \ref{Fig14}(b) and (c) present, respectively, contour plots predicted by the theory for the phase field $z$ and \emph{in-situ} images of the cracks from the experiments at select increasing values of the applied deformation $\lambda=h/H$. The phase field $z$ is shown both from a 3D perspective over the octant (\ref{Oct}) of the specimen in which the simulation is carried out, as well as over its midplane, both in the deformed configuration.

\begin{figure}[t!]
\centering
\centering\includegraphics[width=0.89\linewidth]{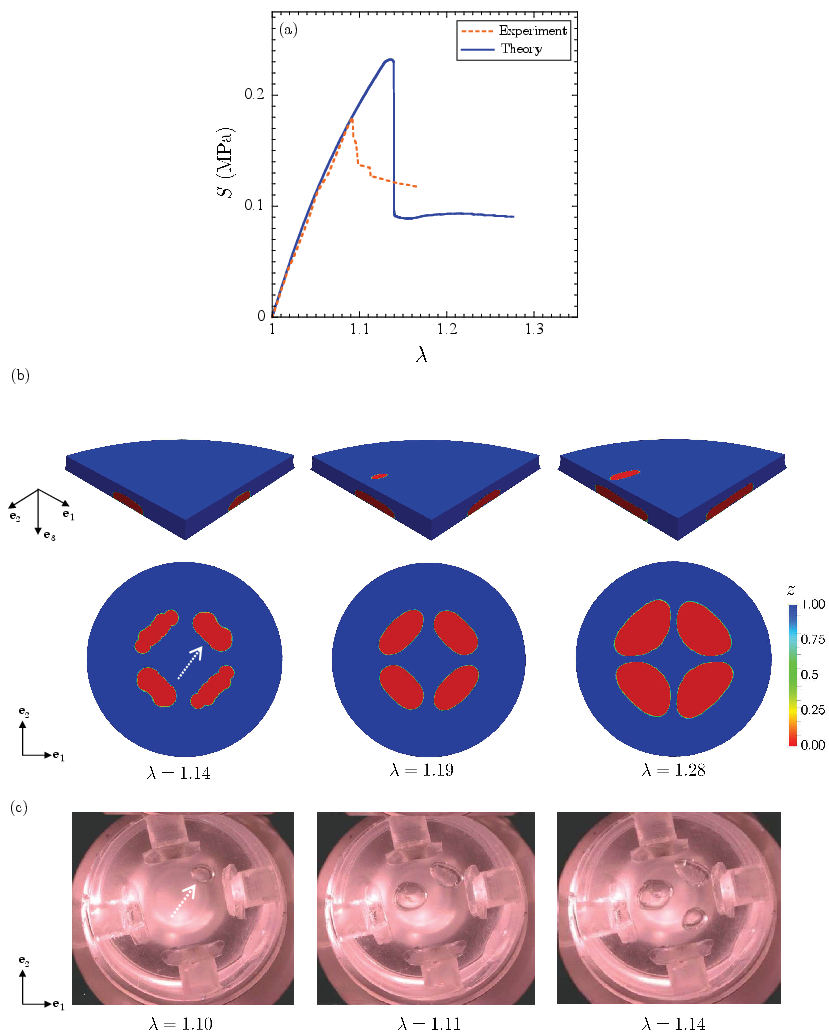}
\caption{\small Comparison between theory and experiment for the thin poker-chip specimen with diameter-to-thickness ratio $D/H=11.3$. (a) The normalized force $S=4P/(\pi D^2)$ as a function of the normalized applied deformation $\lambda=h/H$. (b) Contour plots, over the deformed configuration, of the phase field $z$ at three select applied deformations $\lambda$, as predicted by the theory. (c) Selected sequence of images from the experiment at three different values of $\lambda$. The white arrows in parts (b) and (c) identify the first nucleated crack. \vspace{0.2cm}}\label{Fig15}
\end{figure}

\begin{figure}[t!]
\centering
\centering\includegraphics[width=0.88\linewidth]{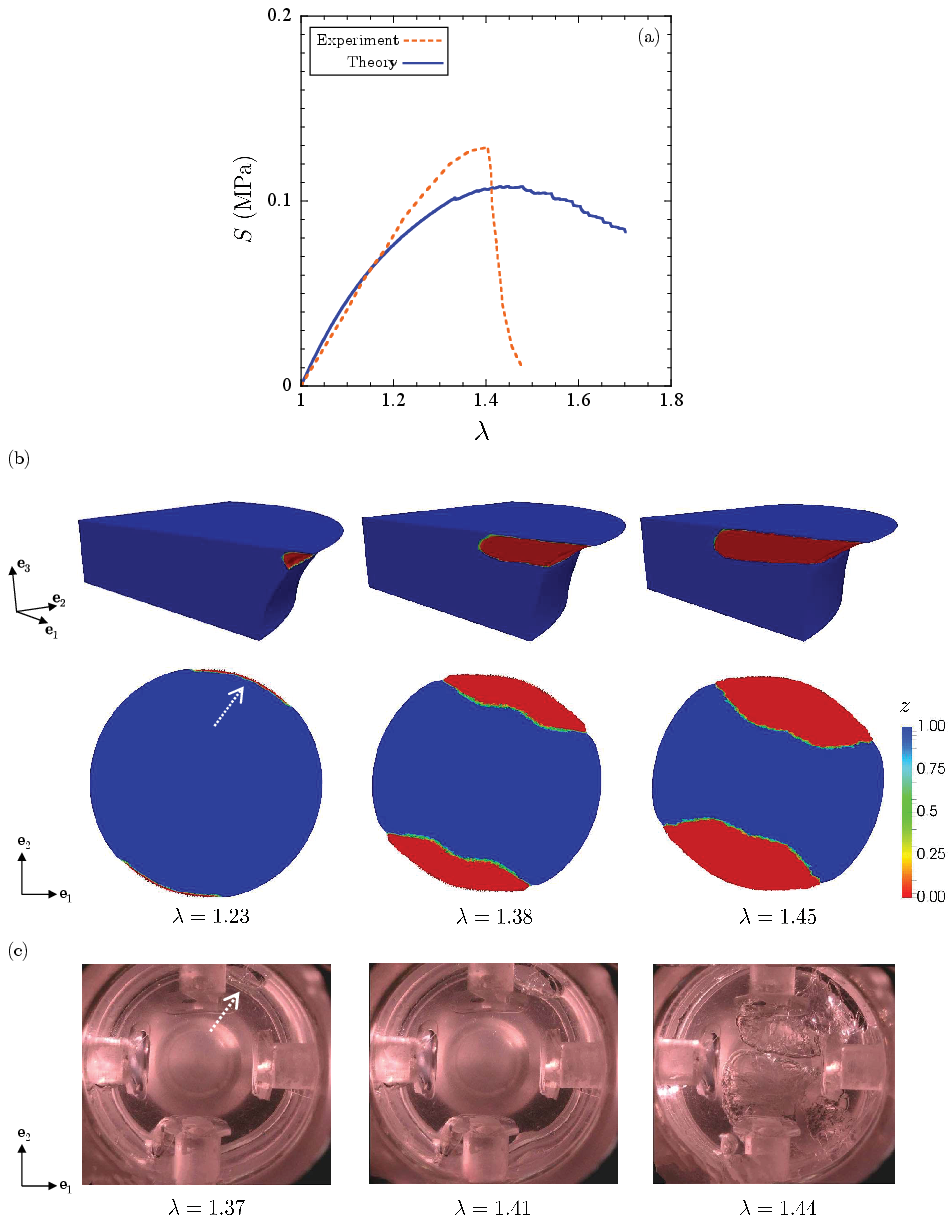}
\caption{\small Comparison between theory and experiment for the thick poker-chip specimen with diameter-to-thickness ratio $D/H=4.7$. (a) The normalized force $S=4P/(\pi D^2)$ as a function of the normalized applied deformation $\lambda=h/H$. (b) Contour plots, over the deformed configuration, of the phase field $z$ at three select applied deformations $\lambda$, as predicted by the theory. (c) Selected sequence of images from the experiment at three different values of $\lambda$. The white arrows in parts (b) and (c) identify the first nucleated crack.}\label{Fig16}
\end{figure}

Figures \ref{Fig15} and \ref{Fig16} present analogous comparisons between the results predicted by the theory with the experimental data for the thin and the thick poker-chip experiments with diameter-to-thickness ratios $D/H=11.3$ and $4.7$; see Figs. 6 and 4 in \citep{GuoRavi23}. The second set of the contour plots for the phase field $z$ are shown \emph{not} over the midplane ($X_3=0$) of the specimens but over the higher lines of latitude $X_3=H/2.21=1.00$ mm and $X_3=H/2.03=2.62$ mm, respectively.

\begin{figure}[t!]
\centering
\centering\includegraphics[width=0.41\linewidth]{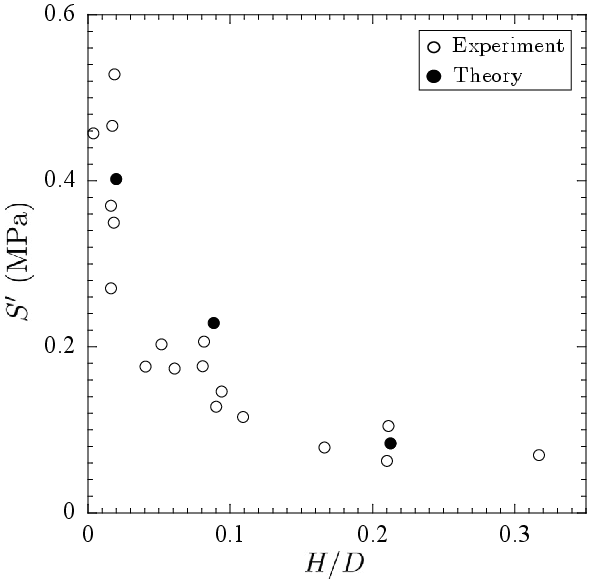}
\caption{\small Comparison between theory and experiment for the critical value $S^\prime$ of the normalized global force $S=4P/(\pi D^2)$ at which the first nucleation of cracks occurs as a function of the normalized thickness $H/D$ of the specimen. For completeness, the figure includes the experimental results for all the tests reported by \cite{GuoRavi23}.}\label{Fig17}
\end{figure}
\begin{remark}
\emph{For the very thin and the thin poker-chip experiments with diameter-to-thickness ratios $D/H=50$ and $11.3$, \cite{GuoRavi23} reported in Figs. 8(a) and 6(a) of their paper the normalized force $S=4P/(\pi D^2)$ \emph{not} as a function of the applied deformation $\lambda=h/H$ but as a function of time $t$. For the thick poker-chip experiment with $D/H=4.7$, they did report the $S$ $vs.$ $\lambda$ data in Fig. 23. In the experimental results that are presented in Figs. \ref{Fig14} and \ref{Fig15}, we have made use of a linear conversion between $t$ and $\lambda$ that leads to results that are consistent with the initial elastic response of the specimens (before the nucleation of any cracks). For the very thin specimen with $D/H=50$, we have used the conversion $\lambda=1.001+\dot{h}_0 t/H$ with $\dot{h}_0/H=2.07\times 10^{-3}$ s$^{-1}$, while for the thin specimen with $D/H=11.3$, we have used the conversion $\lambda=1.01+\dot{h}_0 t/H$ with $\dot{h}_0/H=0.47\times 10^{-3}\,{\rm s}^{-1}$.}
\end{remark}

The main observation from Figs. \ref{Fig14} through \ref{Fig16} is that the predictions generated by the theory are in good qualitative agreement with the experimental results. This is so for where and when the cracks nucleate, as well as for where and when they propagate. The quantitative agreement between the theory and the experiments is also reasonably good, especially taking into account that the simulations are carried out over an octant of the specimen, which overly enforces symmetry, that the values used for the uniaxial tensile and hydrostatic strengths and the critical energy release rate are estimates, and that the experiments exhibit significant stochasticity; see Fig. \ref{Fig17}. 

\paragraph{Where cracks first nucleate} In particular, consistent with the parametric results presented in Section \ref{Sec: 2D results} above for the similar elastomer with hydrostatic-to-tensile strength ratio $\shs/\sts=3$ and critical energy release rate $G_c=75$ N/m, the theoretical results in Figs. \ref{Fig14} and \ref{Fig15} for the two thinner specimens show that the first nucleation of cracks occurs radially away from the centerline of the specimen, in agreement with the experiments. Furthermore, the results in Fig. \ref{Fig16} for the thick specimen show that the nucleation of cracks occurs from the free boundary of the specimen, next to one of the fixtures, also in agreement with the experiment. Again, such locations are drastically different from the ones found in natural rubber because the strength surface of natural rubber is very different from the strength surface of PDMS 30:1.

\paragraph{When cracks first nucleate} To aid the visualization of when the first nucleation of cracks occurs,  Fig. \ref{Fig17} provides a plot of the critical values $S^\prime$ of the normalized global force $S=4P/(\pi D^2)$ at which the first nucleation of cracks occurs in terms of the normalized thickness $H/D$ of the specimen. For completeness and to also illustrate the significant stochasticity observed in the experiments, the figure includes the data for all the tests reported by \cite{GuoRavi23}.

\paragraph{Subsequent crack nucleation and propagation} The theoretical results in Figs. \ref{Fig14} through \ref{Fig16} show that the decrease in the diameter-to-thickness ratio $D/H$ of the specimen leads to a smaller total number of cracks that are nucleated. In other words, in agreement with experiments, the very thin specimen favors additional crack nucleations over the propagation of the initially nucleated cracks, whereas the thick specimen favors the propagation of the initially nucleated cracks over the nucleation of additional cracks. The thin specimen exhibits an intermediate behavior.

\section{Summary and final comments}\label{Sec: Final Comments}

In a nutshell, the results presented in Sections \ref{Sec: strength analysis}, \ref{Sec: 2D results}, and \ref{Sec: Experiments} have established that:
\begin{enumerate}[label=\roman*.,font=\itshape]

\item{The nucleation of internal cracks in poker-chip experiments of synthetic elastomers is dominated by the strength --- in particular, the entire first octant $(s_1>0,s_2>0,s_3>0)$ of the strength surface $\mathcal{F}(\bfS)=0$ --- of the elastomer. That is, internal cracks nucleate in regions where the strength surface of the elastomer has been exceeded, soon after it has been exceeded.}

\item{For a synthetic elastomer whose hydrostatic strength $\shs$ is comparable to or smaller than its uniaxial $\sts$ and biaxial $\sbs$ tensile strengths, the first nucleation of internal cracks occurs around the centerline of the specimen.}

\item{For a synthetic elastomer whose hydrostatic strength $\shs$ is large relative to its uniaxial $\sts$ and biaxial $\sbs$ tensile strengths, the first nucleation of internal cracks occurs radially away from the centerline of the specimen.}

\item{If in addition to featuring a relatively large hydrostatic strength, the elastomer features a relatively small critical energy release rate $G_c$, the first nucleation of cracks occurs from the free boundary of the specimen near one of the fixtures (i.e., the location farthest from the centerline of the specimen). In this case, the nucleated cracks are \emph{not} internal but external (boundary) cracks.}

\item{The propagation of all nucleated (internal and external) cracks is governed by the Griffith competition between the bulk elastic energy of the synthetic elastomer and its constant intrinsic fracture energy. This competition leads to the nucleation of fewer internal cracks in thicker specimens.}
    
\item{The number of cracks nucleated internally is also dependent on the value of the critical energy release rate $G_c$ relative to the values of the hydrostatic strength $\shs$ and the uniaxial tensile strength $\sts$. In particular, fewer internal cracks nucleate for larger values of $\shs$ and $\sts$ relative to $G_c$, as the propagation of previously nucleated cracks is favored over the nucleation of new cracks.}

\end{enumerate}
These findings --- together with those established earlier for natural rubber \citep{KLP21} --- provide a complete description and explanation of the poker-chip experiments of elastomers at large.

The results presented in Section \ref{Sec: Experiments} also add to the list of validation results presented in companion works \citep{KFLP18,KRLP18,KLP20,KBFLP20,KRLP22,KLDLP24}, where the fracture theory of \citet*{KFLP18} has been confronted to experiments of fracture nucleation and propagation in silicone elastomers, titania, graphite, polyurethane, PMMA, alumina, glass, and natural rubber spanning a broad spectrum of specimen geometries and loading conditions. They thus contribute further evidence supporting that the equations (\ref{BVP-y-theory})-(\ref{BVP-z-theory}) provide a complete framework for the description of fracture nucleation and propagation in elastic brittle materials at large under arbitrary (monotonic and quasistatic) loadings.

Finally, in addition to explaining the poker-chip experiments of synthetic elastomers, this work has introduced the new constitutive prescription (\ref{ce-Final}), with (\ref{cehat-Final}) and (\ref{delta-eps-final}), for the driving force $c_{\texttt{e}}(\bfX,t)$ that describes the material strength in the governing equations (\ref{BVP-y-theory})-(\ref{BVP-z-theory}). In contrast to earlier prescriptions, the new prescription (\ref{ce-Final}) is fully explicit. By the same token, the new prescription (\ref{ce-Final}) makes the dependence of the governing equations (\ref{BVP-y-theory})-(\ref{BVP-z-theory}) on the regularization length $\varepsilon$ explicit. \emph{Inter alia}, this should allow us to undertake the task of passing to the limit as $\varepsilon\searrow 0$ in order to determine the corresponding ``sharp'' theory of fracture that the regularized equations (\ref{BVP-y-theory})-(\ref{BVP-z-theory}) represent.

\section*{Acknowledgements}

This work was supported by the National Science Foundation through the Grant CMMI--2132528. This support is gratefully acknowledged.

\section*{Appendix A. Numerical study revealing the constitutive prescription (\ref{delta-eps-final}) for the coefficient $\delta^\varepsilon$}

The results presented in Fig. \ref{Fig7}, as well as additional results for other values of the regularization length $\varepsilon$ and the critical energy release rate $G_c$ not included here, have shown that the coefficient $\delta^\varepsilon$ in the driving force (\ref{ce-Final}) is of the asymptotic form (\ref{delta-eps-0}). In order to determine how the factors $f_{-1}$ and $f_0$ in expression (\ref{delta-eps-0}) depend on the material constants $\mathcal{W}_{\texttt{ts}}$, $\mathcal{W}_{\texttt{hs}}$, $\sts$, and $\shs$, we carry out additional simulations of the ``pure-shear'' fracture test --- still making use of the geometry $H=5$ mm, $L=50$ mm, and $A=10$ mm utilized in Subsection \ref{Sec: delta} in the main body of the text --- varying these constants one at a time while keeping the remaining constants fixed. Specifically, the representative results presented in this appendix pertain to elastomers whose elasticities are described by the stored-energy function (\ref{W-LP}), with 
\begin{equation*}
\mu_2=0\quad {\rm and}\quad  \alpha_1=\alpha_2=1,
\end{equation*}
the critical energy release rate 
\begin{equation*}
G_c=75\, {\rm N/m},
\end{equation*}
and the fifteen different sets of elasticity and strength constants listed in Table \ref{Table2}. All the results pertain to the regularization length 
\begin{equation*}
\varepsilon=0.01\, {\rm mm}
\end{equation*}
and unstructured FE meshes of the uniform size $\texttt{h}/\varepsilon=1/5$.

\begin{table}[H]\centering
\caption{The elasticity and the strength constants used to probe the dependence of the coefficient $\delta^\varepsilon$ on $\mathcal{W}_{\texttt{ts}}$, $\mathcal{W}_{\texttt{hs}}$, $s_{\texttt{ts}}$, and $s_{\texttt{hs}}$.}
\begin{tabular}{cc|cccc}
\toprule
Elasticity constants  &  & Strength constants & &  & \\
\toprule
$\mu_1$ (MPa)   & $\kappa$ (MPa) & $\mathcal{W}_{\texttt{ts}}$ (MPa) & $\mathcal{W}_{\texttt{hs}}$ (MPa)& $s_{\texttt{ts}}$ (MPa) & $s_{\texttt{hs}}$ (MPa)\\
\toprule
$0.015$   & $0.0485$ &  1.8980 & $0.2147$ & 0.24 & 0.36 \\
\midrule
$0.05$  & $0.05$ &  0.5174 & 0.2147 & 0.24 & 0.36 \\
\midrule
$0.08$   & $0.0503$ &  0.2898 & 0.2147 & 0.24 & 0.36 \\
\midrule
$0.15$   & $0.0483$ &  0.1284 & 0.2147 & 0.24 & 0.36 \\
\midrule
$0.25$   & $0.0409$ &  0.0695 & 0.2147 & 0.24 & 0.36 \\
\midrule
$0.35$   & $0.0297$ &  0.0479 & 0.2147 & 0.24 & 0.36 \\
\midrule
\midrule
$0.0801$  & $0.1602$ &  0.2898 & $0.1310$ & 0.24 & 0.36 \\
\midrule
$0.0802$  & $0.8023$ &  0.2898 & 0.0519 & 0.24 & 0.36 \\
\midrule
$0.0803$   & $8.027$ &  0.2898 & 0.0076 & 0.24 & 0.36 \\
\midrule
$0.0803$  & $80.28$ &  0.2898 & 0.0008 & 0.24 & 0.36 \\
\midrule
\midrule
$0.0824$   & $8\times 10^{-5}$ &  0.2898 & $0.1310$ & 0.24 & 0.10 \\
\midrule
$0.0800$   & $0.0800$ &  0.2898 & $0.1310$ & 0.24 & 0.29 \\
\midrule
$0.0802$   & $0.8023$ &  0.2898 & $0.1310$ & 0.24 & 0.63 \\
\midrule
$0.0802$   & $16.05$ &  0.2898 & $0.1310$ & 0.24 & 2.22 \\
\midrule
$0.0803$  & $80.28$ &  0.2898 & $0.1310$ & 0.24 & 4.76 \\
\bottomrule
\end{tabular} \label{Table2}
\end{table}

\begin{figure}[t!]
  \subfigure[]{
   \begin{minipage}[]{0.5\textwidth}
   \centering \includegraphics[width=2.45in]{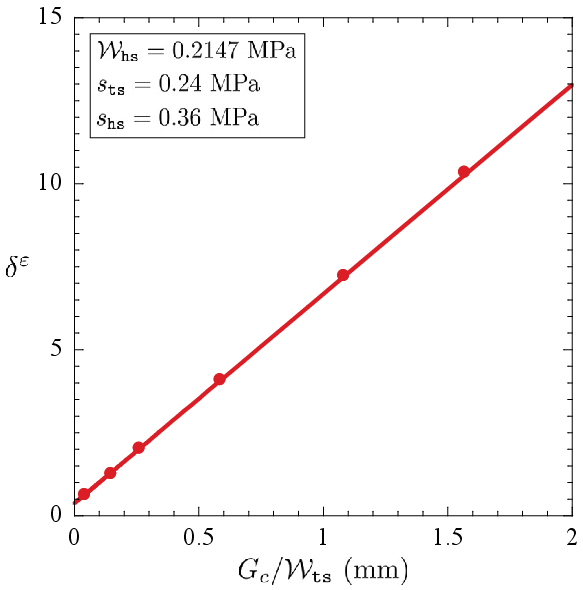}
   \vspace{0.2cm}
   \end{minipage}}
  \subfigure[]{
   \begin{minipage}[]{0.5\textwidth}
   \centering \includegraphics[width=2.5in]{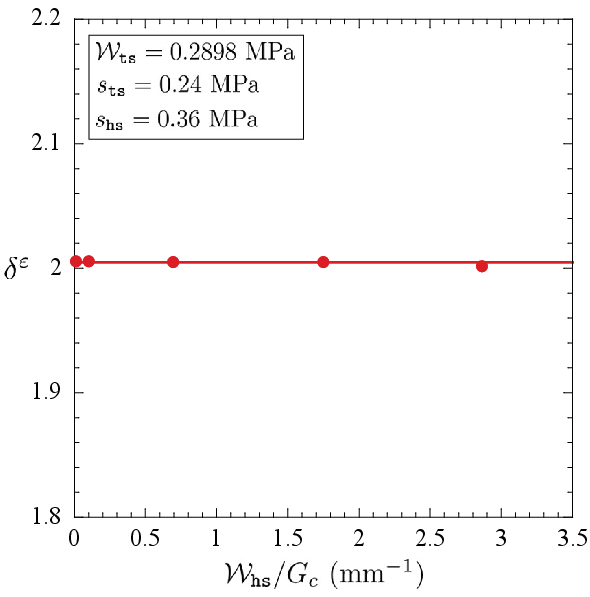}
   \vspace{0.2cm}
   \end{minipage}}
   \caption{Values (solid circles) of the coefficient $\delta^\varepsilon$ determined by matching the theoretical prediction to the Griffith criticality condition (\ref{Griffith-PS}) in a ``pure-shear'' test plotted: (a) as a function of $\mathcal{W}_{\texttt{ts}}$ for fixed $\mathcal{W}_{\texttt{hs}}$, $\sts$, $\shs$ and (b) as a function of $\mathcal{W}_{\texttt{hs}}$ for fixed $\mathcal{W}_{\texttt{ts}}$, $\sts$, $\shs$. For direct comparison, the results generated by the formula (\ref{delta-eps-final-h}) are included (solid lines) in the plots.}\label{Fig_AppAab}
\end{figure}

\begin{figure}[H]
   \centering \includegraphics[width=2.55in]{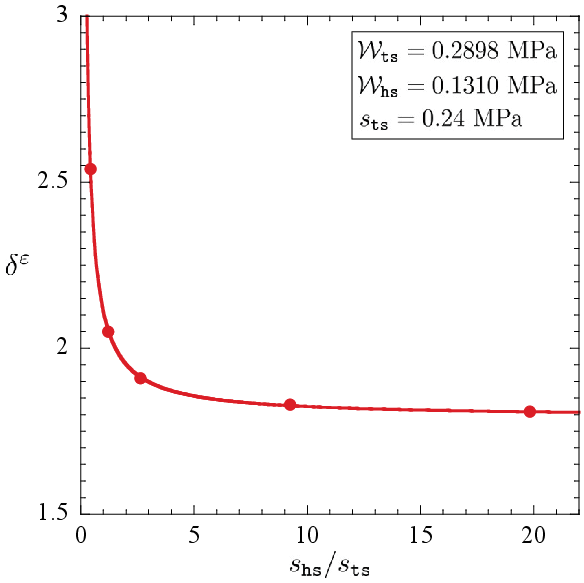}
   \caption{Values (solid circles) of the coefficient $\delta^\varepsilon$ determined by matching the theoretical prediction to the Griffith criticality condition (\ref{Griffith-PS}) in a ``pure-shear'' test plotted as a function of the strength ratio $\shs/\sts$ for fixed $\mathcal{W}_{\texttt{ts}}$, $\mathcal{W}_{\texttt{hs}}$, and $\sts$. For direct comparison, the results generated by the formula (\ref{delta-eps-final-h}) are included (solid lines) in the plot.}\label{Fig_AppAc}
\end{figure}

Figures \ref{Fig_AppAab} and \ref{Fig_AppAc} present the computed values of the coefficient $\delta^\varepsilon$ for which the phase-field theory predicts that the crack in the ``pure-shear'' test starts growing according to (\ref{Griffith-PS}). While Fig. \ref{Fig_AppAab} shows the results as a function of $\mathcal{W}_{\texttt{ts}}$ and $\mathcal{W}_{\texttt{hs}}$, Fig. \ref{Fig_AppAc} shows the results as a function of the strength ratio $\shs/\sts$. To aid the visualization of the entire range of values considered, the results in Figs. \ref{Fig_AppAab}(a) and \ref{Fig_AppAab}(b) are plotted as functions of $G_c/\mathcal{W}_{\texttt{ts}}$ and $\mathcal{W}_{\texttt{hs}}/G_c$, respectively.

%
%
%
%

On the one hand, the results in Figs. \ref{Fig_AppAab}(a) and \ref{Fig_AppAab}(b) show that the coefficient $\delta^\varepsilon$ increases roughly linearly in $1/\mathcal{W}_{\texttt{hs}}$ and is essentially independent of $\mathcal{W}_{\texttt{hs}}$. On the other hand, the results in Fig. \ref{Fig_AppAc} show that $\delta^\varepsilon$ increases monotonically with decreasing strength ratio $\shs/\sts$. Combining these observations and numerous others for different material parameters $\mathcal{W}_{\texttt{ts}}$, $\mathcal{W}_{\texttt{hs}}$, $\sts$, $\shs$ reveal that the expressions (\ref{factors}) provide reasonably accurate approximations for the factors $f_{-1}$ and $f_0$ in (\ref{delta-eps-0}). Corresponding results for different element sizes in the range $\texttt{h}/\varepsilon\in[1/3,1/40]$ further reveal that the formula (\ref{delta-eps-final-h}) provides a reasonably accurate approximation that accounts 
for the discretization error incurred in FE computations based on first-order elements. The results generated by this formula are plotted (solid lines) in Figs.  \ref{Fig_AppAab} and \ref{Fig_AppAc} for direct comparison with the data (solid circles) obtained numerically for $\delta^\varepsilon$.

\section*{Appendix B. A comment on the intrinsic material length scales in the fracture theory (\ref{BVP-y-theory-reg}) and (\ref{BVP-z-theory-reg})}

As noted in Subsection \ref{Sec: Nucleation} in the main body of the text, a key aspect of the fracture theory (\ref{BVP-y-theory-reg})-(\ref{BVP-z-theory-reg}) is that it features (not just one but) a family of intrinsic material length scales.

%
\begin{figure}[b!]
\centering
\includegraphics[width=0.85\linewidth]{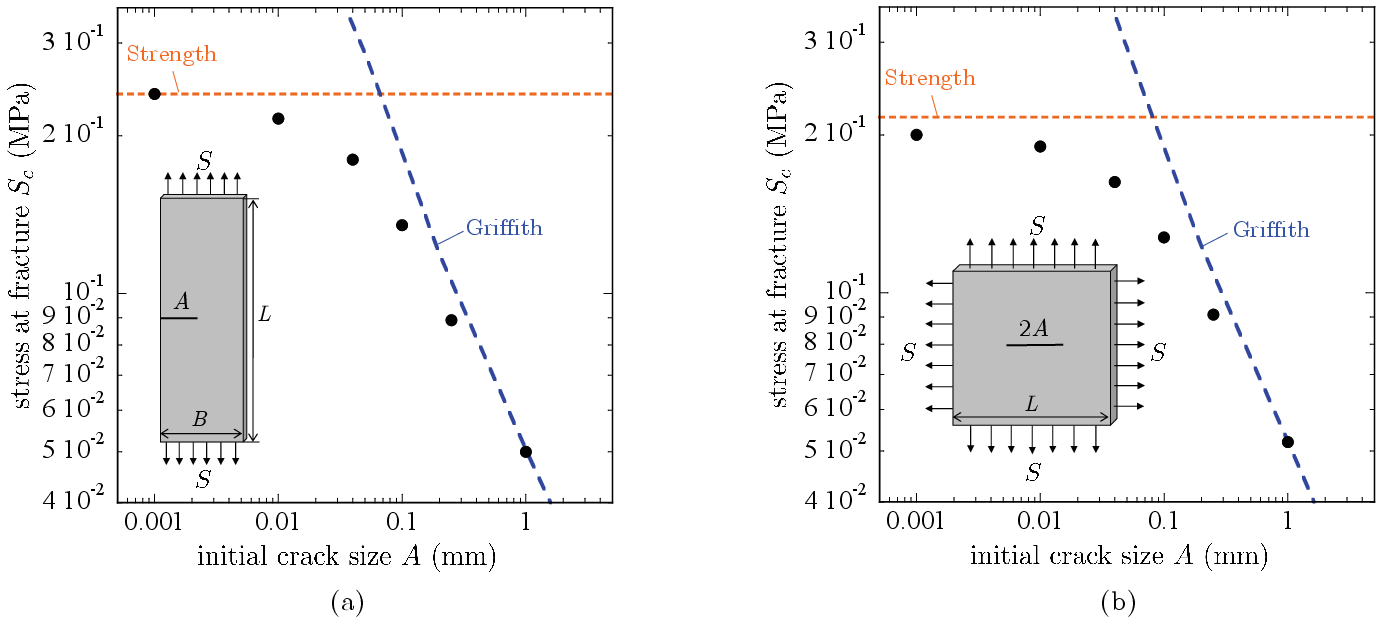}
\caption{{\small Theoretical predictions --- from the simulations of the tests schematically depicted in the insets --- illustrating the transition from Griffith-dominated to strength-dominated nucleation of fracture as the size $A$ of the crack decreases from  large to small. The results show the critical stress $S_c$ at which fracture nucleates from the crack as a function of its size $A$. For direct comparison, the plots include the predictions of nucleation based on the Griffith competition between the elastic and fracture energies (blue dashed line) and based on strength (orange dotted line). The intercepts between these two limiting results roughly identify the material length scales built in the theory corresponding to these two cases.}}\label{Fig_AppB}
\end{figure}
%

To illustrate this key aspect, Fig. \ref{Fig_AppB} presents the predictions from the theory for the critical values $S_c$ of the applied nominal stress $S$ at which fracture nucleates from the pre-existing crack in single-edge crack specimens subjected to uniaxial tension and in center crack specimens subjected to equibiaxial tension; see the insets in Figs. \ref{Fig_AppB}(a) and  \ref{Fig_AppB}(b). The results pertain to thin rectangular plates, of height $L=50$ mm and width $B=10$ mm for the specimens under uniaxial tension and of height $L=50$ mm and width $B=L=50$ mm for the specimens under equibiaxial tension, that are made of an elastomer with stored-energy function (\ref{W-LP}) and the elasticity constants listed in Table \ref{Table1}, Drucker-Prager strength surface (\ref{DP}) with uniaxial tensile and hydrostatic strengths $\sts=0.24$ MPa and $\shs=0.72$ MPa, and critical energy release rate $G_c=75$ N/m. For both problems, it suffices to consider the same value of regularization length $\varepsilon=0.02$ mm. The results are shown as a function of the size $A$ of the pre-existing crack.

Both configurations are standard problems in the literature of fracture of rubber; see, e.g., \citep{Greensmith60,Lindley72,ExtrandGent91,Yeoh02}. The critical values $S_c$ at which \emph{large cracks} start growing are determined by the Griffith criticality condition
\begin{equation}
-\dfrac{\partial \mathcal{U}}{\partial \mathrm{\Gamma}_0}=G_c,\label{Gc-Gen}
\end{equation}
where the left-hand side $-\partial \mathcal{U}/\partial \mathrm{\Gamma}_0$ denotes the change in total deformation energy in the specimen at hand with respect to an added surface area ${\rm d}\mathrm{\Gamma}_0$ to the pre-existing crack $\mathrm{\Gamma}_0$ (i.e., the energy release rate), which can be readily computed via FE; see, e.g., Sections 3 in \citep{SLP23} and \citep{SLP23b}. In the opposite limit of \emph{small cracks}, it is trivial to deduce that
\begin{equation}
S_{c}=\sts\label{S-uni}
\end{equation}
for the specimens under uniaxial tension and
\begin{equation}
S_{c}=s_{\texttt{bs}}=\left(\dfrac{1}{3}+\dfrac{\shs}{\sts}\right)^{-1}\shs\label{S-bi}
\end{equation}
for the specimens under equibiaxial tension. Both sets of limiting results (\ref{Gc-Gen}) and (\ref{S-uni})-(\ref{S-bi}) are plotted in Fig. \ref{Fig_AppB} to aid the discussion.

The results in Fig. \ref{Fig_AppB}(a) show that for crack sizes  $A>0.3$ mm, nucleation of fracture is characterized by the Griffith competition between the elastic and fracture energies of the material (blue dashed line). On the other hand, for crack sizes  $A<0.005$ mm, nucleation of fracture is characterized by the strength of the material (orange dotted line), in this case, its uniaxial tensile strength $s_{\texttt{ts}}=0.24$ MPa. Finally, for crack sizes in the intermediate range  $A\in[0.005,0.3]$ mm, the results  show that nucleation of fracture is characterized by an interpolation between the strength and the Griffith competition between the elastic and fracture energies. The material length scale in this case is about $0.06$ mm.

Similar to Fig. \ref{Fig_AppB}(a), the results in Fig. \ref{Fig_AppB}(b) show that for crack sizes  $A>0.4$ mm, nucleation of fracture is characterized by the Griffith competition between the elastic and fracture energies of the material (blue dashed line), while for crack sizes  $A<0.005$ mm, nucleation of fracture is characterized by the strength of the material (orange dotted line), in this case, its biaxial tensile strength $s_{\texttt{bs}}=0.216$ MPa. For crack sizes in the intermediate range $A\in[0.005,0.4]$ mm, the results  show that nucleation of fracture is characterized by an interpolation between the strength and the Griffith competition between the elastic and fracture energies. The material length scale in this case is about $0.08$ mm.

\bibliographystyle{elsarticle-harv}
\bibliography{References}

\end{document}